\documentclass{aa}  

\usepackage{natbib}
\usepackage{graphicx}
\usepackage{xcolor}
\usepackage{hyperref}

\usepackage{txfonts}

\newcommand{\kms}{km\,s$^{-1}$}
\newcommand{\ms}{m\,s$^{-1}$}
\newcommand{\masy}{mas\,y$^{-1}$}
\newcommand{\mpl}{\mbox{M$_{p}$}}
\newcommand{\rpl}{\mbox{R$_{p}$}}
\newcommand{\mstar}{\mbox{M$_{s}$}}
\newcommand{\rstar}{\mbox{R$_{s}$}}
\newcommand{\mjup}{\mbox{M$_{\rm J}$}}

\newcommand{\msun}{\mbox{M$_{\odot}$}}
\newcommand{\rsun}{\mbox{R$_{\odot}$}}
\newcommand{\rearth}{R$_{\oplus}$}
\newcommand{\mearth}{M$_{\oplus}$}
\newcommand{\gccc}{g\,cm$^{-3}$}

\newcommand{\teff}{$T_{\rm eff}$}
\newcommand{\logg}{$\log g$}
\newcommand{\met}{\mbox{[Fe/H]}}
\newcommand{\age}{\mbox{t$_{s}$}}
\newcommand{\lstar}{\mbox{L$_{s}$}}
\newcommand{\Nstar}{TOI-333}

\newcommand{\Nplanet}{TOI-333b}

\bibpunct{(}{)}{;}{a}{}{,} 

\begin{document}

   \title{\Nplanet: A Neptune Desert planet around a F7V star}
   \authorrunning{Alves et al.}


   \author{Douglas R. Alves\thanks{E-mail: douglasalvesastro12@gmail.com; dalves@das.uchile.cl}\fnmsep
          \inst{1,2}
          James S. Jenkins
          \inst{3,2},
          Jos\'e I. Vin\'es
          \inst{4},
          Maximilano Moyano
          \inst{4},
          David R. Anderson
          \inst{4},
          Christian Magliano
          \inst{5},
          Giovanni Covone
          \inst{5},
          Keivan G.\ Stassun
          \inst{6},
          Abderahmane Soubkiou
          \inst{7},
          Edward Gillen
          \inst{8},
          Matthew P. Battley
          \inst{8},
          Alexander Hughes
          \inst{8},
          David J. Armstrong
          \inst{9,10},
          Suman Saha
          \inst{3,2},
          Faith Hawthorn
          \inst{9,10},
          Peter J. Wheatley
          \inst{9,10},
          Karen A.\ Collins
          \inst{11},
          Richard P. Schwarz
          \inst{11},
          Gregor Srdoc
          \inst{12},
          Ioannis Apergis 
          \inst{9,10},
          Tafadzwa Zivave
          \inst{9,10},
          Monika Lendl
          \inst{13},
          Benjamin M. Tofflemire
          \inst{14},
          John~P.~Doty
          \inst{15},
          Christina Hedges
          \inst{16},
          Ismael Mireles
          \inst{17},
          Matthew R. Burleigh
          \inst{18},
          Alicia Kendall
          \inst{18},
          George T. Harvey
          \inst{18},
          Michael R. Goad
          \inst{18},
          Sarah L. Casewell
          \inst{18},
          Troy Edkins
          \inst{18}
          }

   \institute{Departamento de Astronom\'ia, Universidad de Chile, Casilla 36-D, Santiago, Chile
   \and
   Centro de Astrof\'isica y Tecnolog\'ias Afines (CATA), Casilla 36-D, Santiago, Chile
   \and
   Instituto de Estudios Astrof\'isicos, Universidad Diego Portales, Av. Ej\'ercito 441, Santiago, Chile
   \and
   Instituto de Astronomía, Universidad Católica del Norte, Angamos 0610, Antofagasta 1270709, Chile
   \and
   Department of Physics “Ettore Pancini”, University of Naples Federico II, Naples, Italy
   \and
   Department of Physics and Astronomy, Vanderbilt University, Nashville, TN 37235, USA
   \and
   Astrobiology Research Unit, Université de Liège, Allée du 6 août 19, Liège, 4000, Belgium
   \and
   Astronomy Unit, Queen Mary University of London, Mile End Road, London E1 4NS, UK
   \and
   Department of Physics, University of Warwick, Gibbet Hill Road, Coventry CV4 7AL, UK
   \and
   Centre for Exoplanets and Habitability, University of Warwick, Gibbet Hill Road, Coventry CV4 7AL, UK
   \and
   Center for Astrophysics \textbar \ Harvard \& Smithsonian, 60 Garden Street, Cambridge, MA 02138, USA
   \and
   Kotizarovci Observatory, Sarsoni 90, 51216 Viskovo, Croatia
   \and
   Observatoire astronomique de l'Université de Genève, Chemin Pegasi 51, 1290 Versoix, Switzerland
   \and
   SETI Institute, Mountain View, CA 94043 USA/NASA Ames Research Center, Moffett Field, CA 94035 USA
   \and
   Noqsi Aerospace Ltd., 15 Blanchard Avenue, Billerica, MA 01821, USA
   \and
   NASA Goddard Space Flight Center, 8800 Greenbelt Rd, Greenbelt, MD 20771, USA
   \and
   Department of Physics and Astronomy, University of New Mexico, 210 Yale Blvd NE, Albuquerque, NM 87106, USA
   \and
   School of Physics and Astronomy, University of Leicester, Leicester LE1 7RH, UK
             }

   \date{Received month day, year; accepted month day, year}

\abstract{Observations have shown that planets similar to Neptune are rarely found orbiting Sun-like stars with periods up to $\sim$4 days, defining the so-called Neptune desert region. Therefore, the detection of each individual planet in this region holds a high value, providing detailed insights into how such a population came to form and evolve. Here we report the detection of \Nplanet\,, a Neptune desert planet with a mass, radius, and bulk density of 20.1 $\pm$ 2.4 M$_{\oplus}$, 4.26 $\pm$ 0.11 R$_{\oplus}$, and 1.42 $\pm$ 0.21 \gccc, respectively. The planet orbits a F7V star every 3.78 d, whose mass, radius and effective temperature are of 1.2 $\pm$ 0.1 \msun, 1.10 $\pm$ 0.03 \rsun, and 6241$^{+73}_{-62}$ K, respectively. \Nplanet\, is likely younger than 1 Gyr, which is supported by the presence of the doublet Li line around 6707.856 $\AA$ and its comparison to Li abundances in open clusters with well constrained ages. The planet is expected to host only 8.5$^{+10.9}_{-8.3}\%$ gas-to-core mass ratio for a H/He envelope. On the other hand, irradiated ocean world models predict 20$^{+11}_{-10}\%$ H$_2$O mass fraction with a core fraction of 35$^{+20}_{-23}\%$. Therefore, we expect that \Nplanet\,internal composition may be dominated by a pure rocky composition with almost no H/He envelope, or a rocky world with almost equal mass fraction of water. Finally, \Nplanet\,is more massive and larger than 77$\%$ and 82$\%$ of its Neptune desert counterparts, respectively, while its host ranks among the hottest known for Neptune Desert planets, making this system a unique laboratory to study the evolution of such planets around hot stars.}
 


   \maketitle
%

\section{Introduction}
Shortly after the launch of the \textit{Kepler} Space Telescope \citep{borucki2010kepler}, the detection of thousands of exoplanets and planet candidates transformed our understanding of the planetary population in the Galaxy. It quickly became evident that the most common types of planets are super-Earths and sub-Neptunes, which orbit approximately 30\% of Sun-like stars \citep{fressin2013false, mulders2018exoplanet}. The \textit{Kepler} mission also revealed the existence of the so-called Neptune desert, an area in the period-radius-mass parameter space that shows a significant dearth of planetary systems \citep{szabo2011short,mazeh2016dearth, castro2024mapping}. The region extends to orbital periods of $\sim$4 days, planetary radii between roughly 2 and 10~R$_\oplus$ and masses about 0.03 - 0.1\mjup. Additionally, the Neptune desert (ND) is further corroborated by the scarce number of detections from subsequent space- \citep[e.g., TESS;][]{ricker2015transiting} and ground-based missions \citep[e.g., NGTS;][]{wheatley2018next}, yet a few outstanding discoveries have been detected. For example, LTT9779b \citep{jenkins2020ultrahot}, an ultra-hot planet where recent studies \citep{hoyer2023extremely,reyes2025closer, saha2025high} indicate the presence of a likely metal-rich atmosphere and silicate clouds, TOI-824 \citep{burt2020toi} a nearby planet twice as dense as Neptune, and TOI-849b \citep{armstrong2020remnant}, an exposed core of what might have been a giant planet.

The ND population has been steadily growing, with each new detection proving particularly valuable. This is especially true for transiting planets, where key parameters (e.g., radii, densities, secondary eclipses) can be measured. Thus, the study of transiting ND planets enable robust statistical analyses aimed at investigating the likely origins of the ND and understanding its subsequent evolution. For instance, it has been shown that short-period NDs have envelope mass fractions approaching zero, with their host stars also being more metal-rich \citep{des2007diagram,doyle2025exploring,vissapragada2025hottest}. This may hint for close-in NDs being the remnants of gas giants, thus indicating that distinct evolutionary pathways may be at play as a function of orbital period but also the host metallicity. Moreover, planet radius and mass distributions for ND planets steeply rise toward smaller, lower-mass planets, following well-established power-law trends \cite{lopez2012effects,petigura2013plateau}. Consequently, larger Neptunes with radii between 4 and 6~R$_\oplus$ are relatively rare, orbiting only about $\sim$3\% of stars. Additionally, large ultrashort period (USP) Neptunes (R~$\sim$~2-6~R$_\oplus$) are among the rarest planets detected, where \textit{Kepler} found virtually none.

ND planets are expected to retain substantial atmospheres overlying rocky or icy cores, making them crucial benchmarks for studying the physics and chemistry of planets in the non–gas giant regime. In particular, USP and ultra-hot Neptunes have atmospheres exceeding 2000~K, where intense irradiation drives a rich mix of neutral and ionized species along with exotic cloud formations \citep{crossfield2020phase,dragomir2020spitzer}. These extreme environments offer unique laboratories for probing atmospheric chemistry under conditions that cannot be reproduced elsewhere. Additionally, the upper atmospheres of these planets can reach escape velocity, leading to strong atmospheric outflows. As a result, they serve as prime targets for investigating atmospheric mass-loss processes \citep[e.g.,][]{mansfield2018detection}.

Although the exact origin of the ND is still unknown, our current leading description involves a combination of tidal migration and photoevaporation \citep{lopez2013role,owen2017evaporation}, where recent theoretical and observational studies argue that photoevaporation alone may not be strong enough to shape the desert’s upper boundary. Instead, tidal migration and tidal disruption appear to be more plausible mechanisms \citep{owen2018photoevaporation,vissapragada2022maximum}. Additionally, Roche lobe overflow (RLO) is likely to play a significant role in stripping massive envelopes from planets that are extremely close to their host stars, with orbital periods of two days or less \citep{valsecchi2015tidally, jackson2016tidal}. Therefore, the detection of ND planets and their subsequent follow-up are key to provide insights into the physical processes responsible for creating and maintaining this region.

We report the discovery of \Nplanet, a short-period planet located in the ND, with a mass of 20.1 $\pm$ 2.4 M$_{\oplus}$, radius of 4.26 $\pm$ 0.11 R$_{\oplus}$, and bulk density of 1.42 $\pm$ 0.21 \gccc. The planet orbits a F7V-type star every 3.78 days. The host star has a mass of 1.2 $\pm$ 0.1 \msun, a radius of 1.10 $\pm$ 0.03 \rsun, and an effective temperature of 6241$^{+73}_{-62}$ K.
In Section \ref{sec:obs}, we describe the multi-instrument observations that led to the planet's detection. Section~\ref{sec:analysis} outlines the photometric validation process, along with the derivation of stellar properties and planetary parameters. Sections \ref{sec:discussion} and \ref{sec:conc} present the discussion and conclusions, respectively.
\section{Observations}

Here we describe the photometric and spectroscopic time series acquisition and data reductions that led to the discovery of \Nplanet. Table \ref{tab:photometry} and \ref{tab:RVs-table} show a portion of the photometry and radial velocity (RV) for guidance. The complete dataset is made available through the online supplementary material.
\label{sec:obs}
%
%
%

\subsection{TESS Photometry}
\label{sub:tessphot}
\Nstar\ was first observed by the Transiting Exoplanet Survey Satellite \citep[TESS;][]{TESS} in Sectors 02, 29, 69 with cadences of 30, 2, and 2 minutes, respectively. The data were acquired from the Mikulski Archive for Space Telescopes (MAST), using the \texttt{lightkurve} package \citep{lightkurve}. The image data were reduced and analysed by the Science Processing Operations Center at NASA Ames Research Center. We opted to use the PDC$\_$SAP \citep{2012PASP..124.1000S,Stumpe2012,Stumpe2014} photometric time series from the Science Processing Operations Center \citep[SPOC;][]{jenkins2016tess, caldwell2020tess}, which are reduced and analysed at NASA Ames Research Center, due to their overall lower scatter compared to other pipelines. Nonetheless, we inspected the data from the Quick Look Pipeline \citep[QLP;][]{kunimoto2021quick}, confirming the transit events there as well.

The transit signal was initially identified in Full Frame Image (FFI) data by the Quick Look Pipeline (QLP) at MIT \citep{2020RNAAS...4..204H, 2020RNAAS...4..206H}. The TESS Science Office (TSO) subsequently reviewed the vetting products and issued an alert on 20 December 2018 \citep{guerrero:TOIs2021ApJS}. We then performed an independent transit search using the Transit Least Squares (TLS) algorithm \citep{hippke2019optimized}, detecting a total of 18 events. A prominent signal was found at a period of 3.785 days, prompting a validation process to determine whether the signal originated from a transiting hot Neptune or was instead a false positive.

As part of our photometric vetting, we examined the transit morphology for signs of odd-even depth differences or V-shaped profiles indicative of background eclipsing binaries. Additionally, we note that the SPOC pipeline provides centroid diagnostics that assess the location of the transit signal relative to the nominal TIC host \citep{Twicken:DVdiagnostics2018}. For this candidate, the offset was measured to be 1.1$\pm$2.8 arcseconds, consistent with the signal originating from TIC 224245334 as shown in Fig. \ref{fig:tesscutfile}.

\begin{figure*}
    \includegraphics[width=0.7\columnwidth,angle=0]{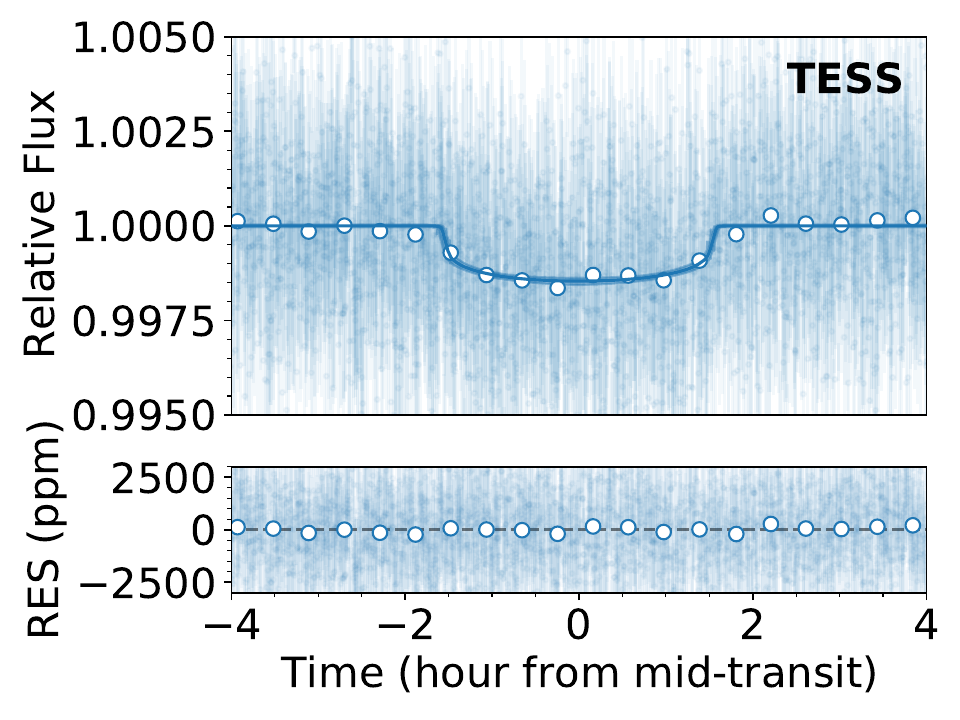}
    \includegraphics[width=0.7\columnwidth,angle=0]{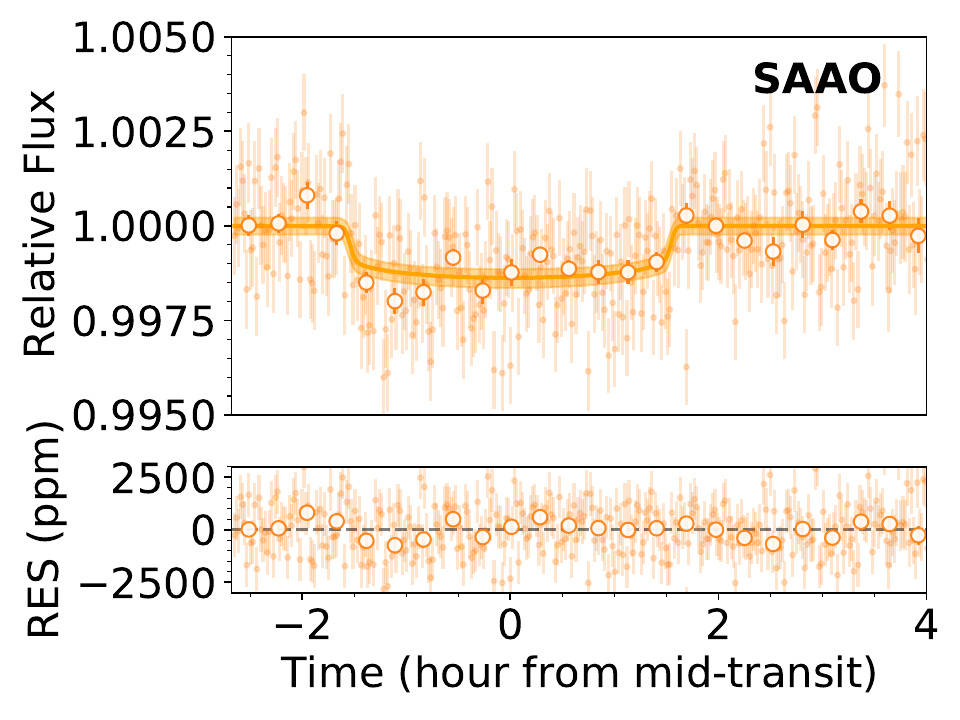}
    \includegraphics[width=0.7\columnwidth,angle=0]{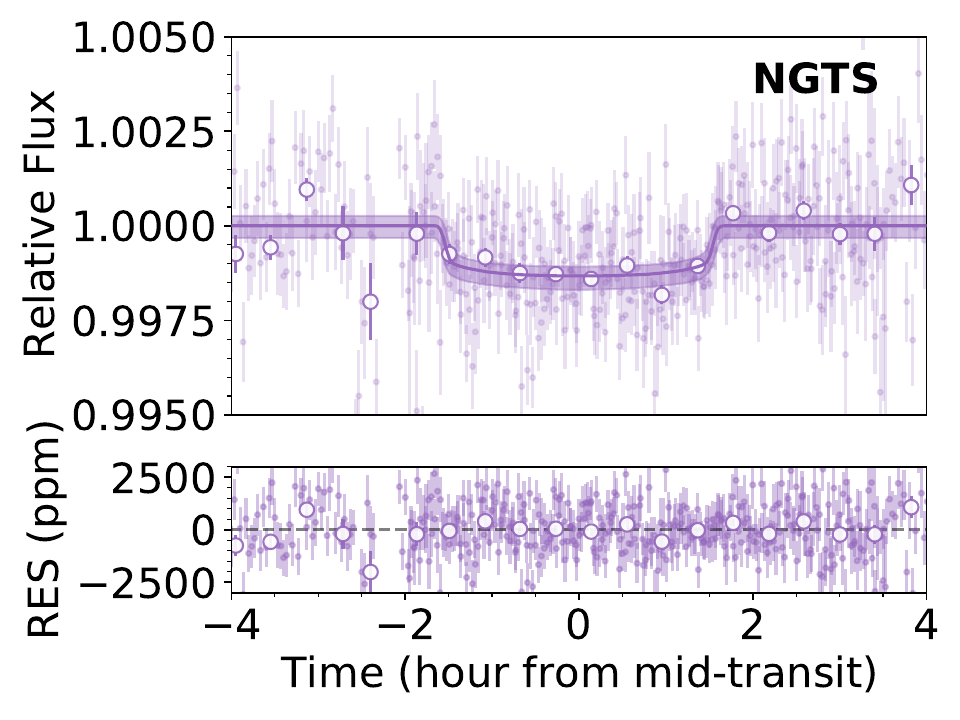}
    \caption{\textbf{Left}: TESS detrended lightcurve phase-folded to the best-fitting period listed in Table \ref{tab:planet} and zoomed to show the transit event. Blue and white circles correspond to modelled photometric data and binned data with the associated photon noise error. The blue line and shaded region show the median transit model and its 1-$\sigma$ confidence interval. \textbf{Centre}: Same as left figure for the LCOGT-SAAO telescope. \textbf{Right}: Same as left figure for the NGTS mission.
    \textbf{Bottom}: residuals to the best fit model.}
    \label{fig:transitModels}
\end{figure*}

Fig. \ref{fig:transitModels} (left) show the detrended phase-folded lightcurves well as the best-fitting transit model derived in $\S$ \ref{sub:globalmodelling}.

\subsection{LCOGT Ground-based Photometry\label{subsec:lcogt}}

We observed a \Nstar\,full transit window on UTC 2021 September 10 in Sloan $i'$ band from the Las Cumbres Observatory Global Telescope (LCOGT) \citep{Brown:2013} 1.0\.m network node at South Africa Astronomical Observatory near Sutherland, South Africa (SAAO). The 1\,m telescope is equipped with a $4096\times4096$ SINISTRO camera having an image scale of $0\farcs389$ per pixel, resulting in a $26\arcmin\times26\arcmin$ field of view. The images were calibrated by the standard LCOGT {\tt BANZAI} pipeline \citep{McCully:2018} and differential photometric data were extracted using {\tt AstroImageJ} \citep{Collins:2017}. We used a circular $7\farcs8$ photometric target star aperture that excluded all flux from the nearest known star (Gaia DR3 6535750174975363712), which is 22$\arcsec$ north of \Nstar. We detected the transit in the target star photometric aperture, which confirms that the TESS detected event is indeed occurring in TOI-333.

Fig. \ref{fig:transitModels} (centre) shows the phase-folded LCOGT follow-up photometry from September 10, 2021 with the best-fitting transit model found from the global modelling in $\S$ \ref{sub:globalmodelling}. We also observed \Nstar\,on six prior epochs (from August 2019 through August 2021), but the transit event was ruled out. After the ephemeris was revised from the TESS sector 69 data, the six earlier observation windows were determined to be out of transit.
\subsection{NGTS Photometry}
\label{sub:ngtsphot}
The Next Generation Transit Survey \citep[NGTS;][]{wheatley2018next} consists of 12 telescopes operating at ESO’s Paranal Observatory in Chile, designed to detect new transiting planetary systems. Each telescope has a 0.2 m aperture and an individual field of view of 8 deg$^2$, resulting in a combined wide-field coverage of 96 deg$^2$. The detectors feature a 2K $\times$ 2K pixel array with 13.5 $\mu$m pixels, corresponding to an on-sky resolution of 4.97 arcseconds. These high-sensitivity detectors operate over a wavelength range of 520$-$890 nm. The setup enables 150 ppm photometric precision for bright stars ($V <$10 mag) in multi-camera mode, while single-telescope observations at a 30-minute cadence achieve a precision of 400 ppm \citep{bayliss2022high}. The mission has been responsible to the detection of 28 planets thus far, from giant hot Jupiters like NGTS-6 \citep{vines2019ngts} and NGTS-21  \citep{alves2022ngts} down to sub-Neptune-sized planet like NGTS-4 \citep{west2019ngts}. Within NGTS treasure-trove are also the discovery of a remarkably young HJ with an age of $\sim$ 30 Myr \citep[NGTS-33b;][]{alves2025ngts}.

TOI-333 follow-up with NGTS occurred during 2024 on September 1, September 19, and October 27, where 8, 6, and 6 telescopes were employed, respectively. A continuum and two transits were captured, with a total of 728 images acquired, with having an exposure time of 10 seconds per frame. Aperture photometry extraction was carried out with the CASUTools\footnote{\url{http://casu.ast.cam.ac.uk/surveys-projects/software-release}} package, and nightly trends such as atmospheric extinction were corrected for with an adapted version of the SysRem algorithm \citep{Tamuz2005}.

Fig.~\ref{fig:transitModels} (right) shows the NGTS detection lightcurve wrapped around the best-fitting period $3.785257 \pm 0.000003$\,d computed from the global modelling ($\S$ \ref{sub:globalmodelling}). For a detailed description of the NGTS mission, data reduction, and acquisition, we refer the reader to \citet{wheatley2018next}.





\subsection{HARPS spectroscopy}
\label{sub:harps-specostrocy}
\Nstar\,has been part of our programme to detect and characterise ND planets, where 37 high-resolution Echelle spectra were obtained during UT 2022-07-02 through 2024-08-14 under the program IDs 109.2374 and 113.26GX (PI: JENKINS) and 112.25QD (PI: ALVES), with the HARPS spectrograph on the ESO 3.6 m \citep{2003Msngr.114...20M} telescope at the La Silla Observatory in Chile. A total of 21.42 hours have been dedicated to \Nstar\,with exposure times of 1800-2100 seconds depending on weather and seeing conditions. The high accuracy mode (HAM) was used, achieving a typical signal-to-noise (SNR) of 25 per pixel at 6500 \AA\,and an RV precision of $\sim8$ \ms. The standard HARPS pipeline \citep{lovis2007new} was used to compute the RVs, where we opted for a G2 binary mask given that it is the most suitable for \Nstar\,, a late F-type star. Fig. \ref{fig:rvsphot} presents the HARPS modelled RVs in light brown, while the periodogram of the RV residuals is shown in Fig. \ref{fig:RES-LS} in the appendix. No statistically significant signal is found above the 10$\%$ false alarm probability (FAP).

\begin{figure}
	\includegraphics[width=\columnwidth]{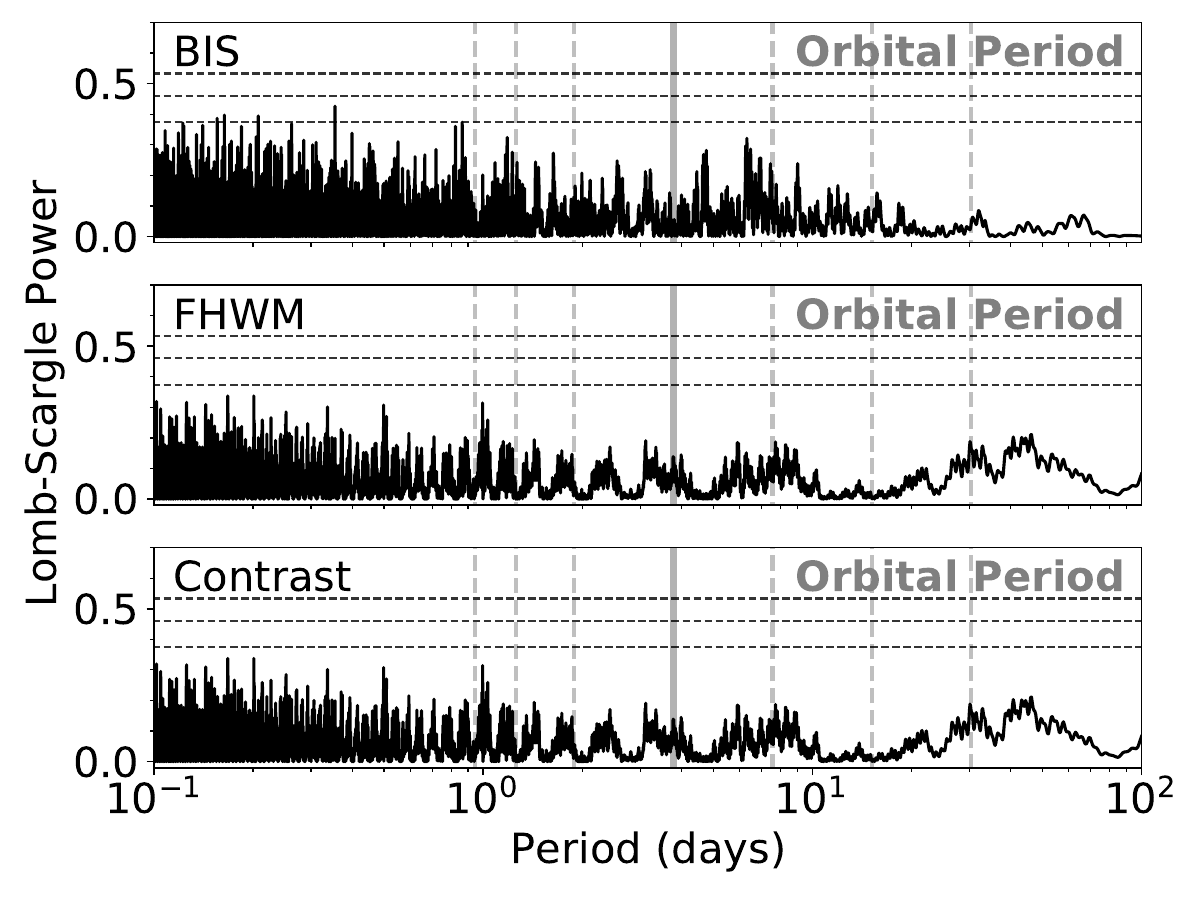}
    \caption{Periodogram of the line bisectors (top panel), CCF-FWHM (centre panel) and contrast (bottom panel) for HARPS (in black). The planet’s orbital period is highlighted by the grey vertical line with dashed lines showing the harmonics at 1/8, 1/4, 1/2, 2, and 3 from left to right, respectively. Top to bottom dashed black lines represent the FAP at 0.1$\%$ 1$\%$ and 10$\%$, respectively.}
    \label{fig:diagnosis}
\end{figure}

\begin{figure}
	\includegraphics[width=\columnwidth]{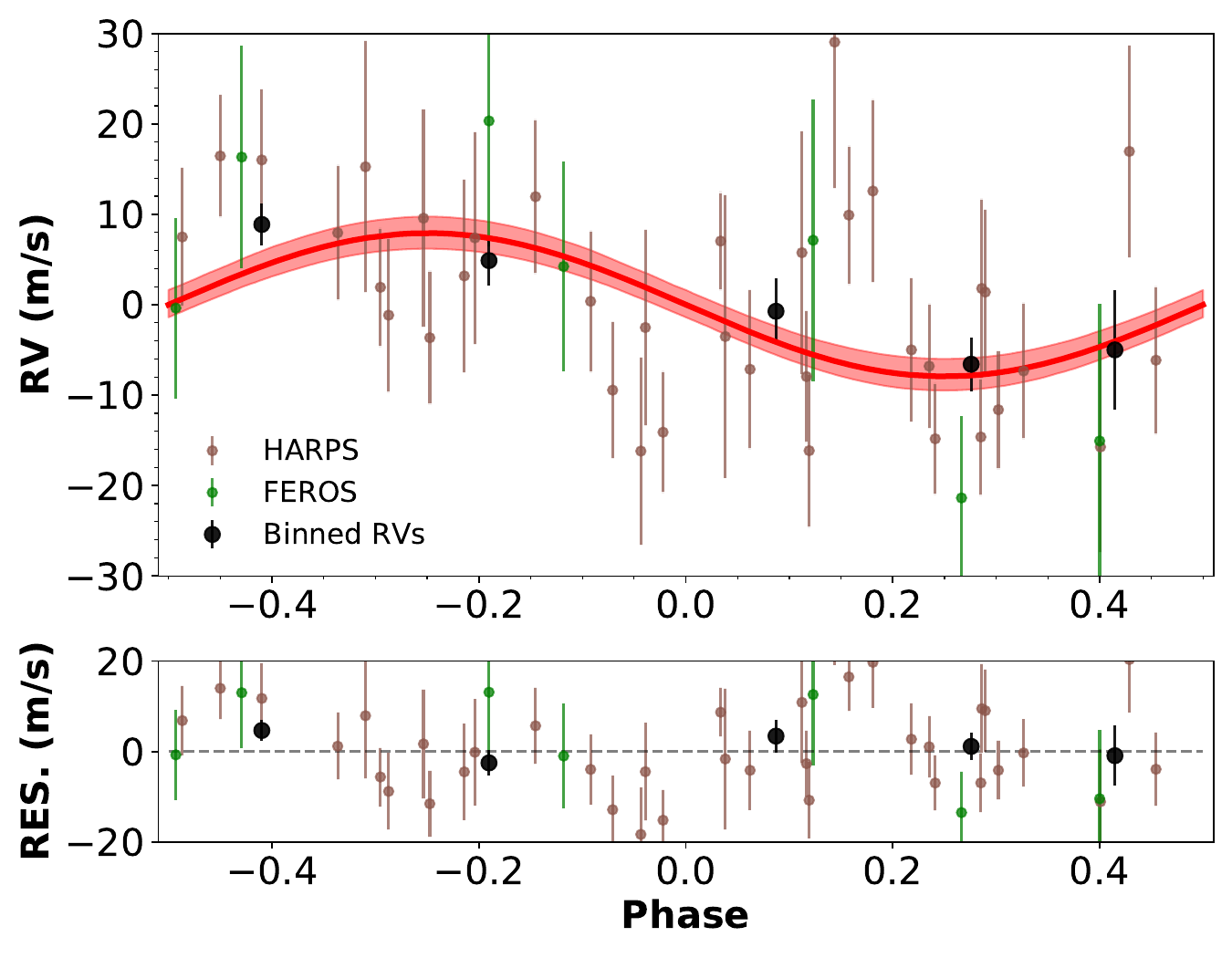}
    \caption{\textbf{Top}: RV phase-folded to the best-fitting period listed in Table \ref{tab:planet}. RV data is colour-coded in brown, and green for HARPS and FEROS, respectively, with black circles representing the binned RVs. The red curve and light red shaded region shows the Keplerian model and its 1-$\sigma$ confidence interval. \textbf{Bottom}: residuals to the best fit model.}
    \label{fig:rvsphot}
\end{figure}

\begin{figure}
	\includegraphics[width=\columnwidth]{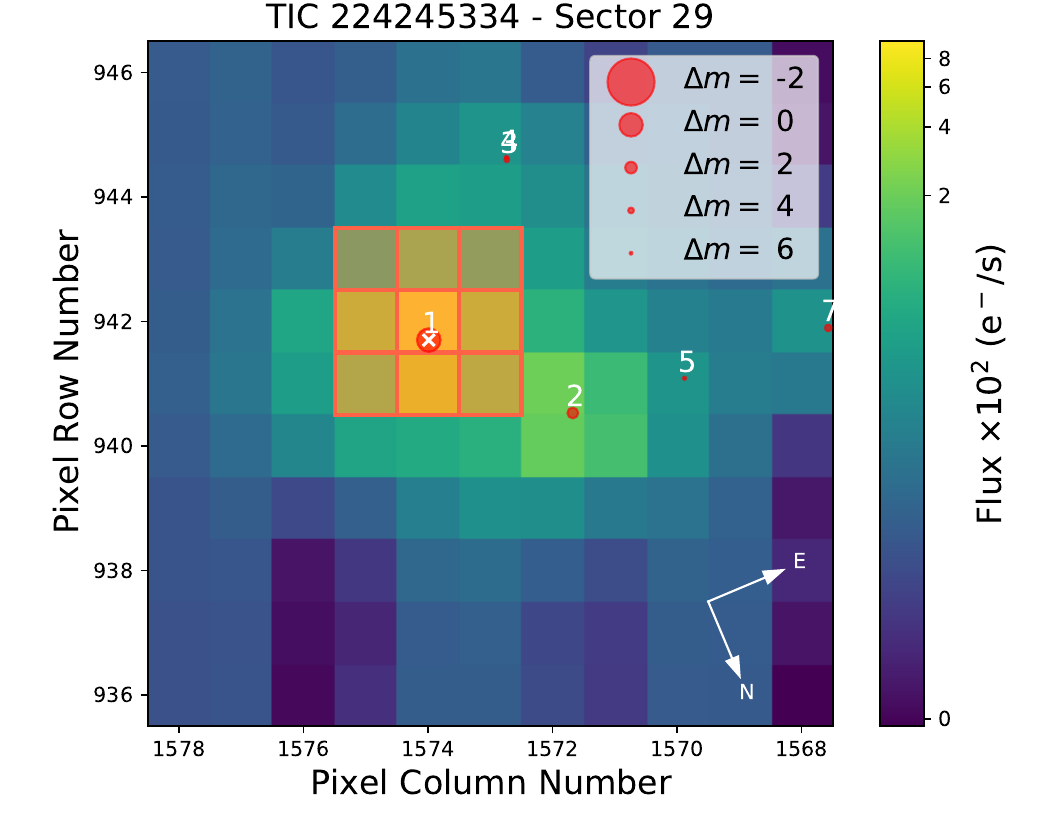}
    \caption{TESS sector 29 full-frame image cutout (11 x 11 pixels) generated with the \texttt{tpfplotter} script described in \citet{aller2020planetary}. \Nplanet\ is shown in the centre labeled number 1, followed by GAIA DR3 6535747602288702720 (number 2), a G = 14.11 mag, and 51" away from the planet. The star did not contribute significant flux to the aperture, thus negligible dilution was observed in TESS.}
    \label{fig:tesscutfile}
\end{figure}

We performed a spectral line diagnostics analysis to assess whether stellar activity could be affecting the RV signals. Specifically, we searched for periodicities in the generalized Lomb-Scargle periodograms of the bisector velocity span (BIS), the full width at half maximum (FWHM), and the contrast of the cross-correlation function (CCF), all derived using the HARPS DRS pipeline, yet no significant signal, particularly close to the planet's period has been found, as shown in Fig. \ref{fig:diagnosis}. Finally, we computed the Pearson r coefficient to investigate the correlation between the RVs, against BIS, FWHM-CCF and contrast as correlations would indicate instrumental and/or stellar effects may be impacting the observed spectral lines. We found weak correlations as a function of BIS, FWHM-CCF and contrast, with values of 0.07, 0.21, and -0.4, respectively.
\subsection{FEROS spectroscopy}
\label{sub:feros}
We obtained 7 high-resolution echelle spectra using the FEROS spectrograph on the MPG/ESO 2.2-m \citep{kaufer1999commissioning} telescope at the La Silla Observatory in Chile. The observations ran from UT 2023-09-16 through 2023-09-28 under the program ID 0111.A-9019(A) (PI: MOYANO), with exposure times of 1200-1800 seconds, depending on weather conditions. The automated {\sc ceres} pipeline \citep{brahm2017ceres} was used for data reduction, where all steps from optimal spectral extraction, wavelength calibration, instrumental drift correction, and continuum normalization were performed. Finally, {\sc ceres} calculated the RVs using the CCF method using a G2 mask, leading to a typical SNR of 21 and RV precision of 14 \ms.
\subsection{Speckle imaging}
Stellar companions or background eclipsing binaries could cause periodic dips in the photometric time series. To rule out such events and provide further assurance that the origin of the transit signal occurs on \Nstar, we performed high-angular resolution imaging with the 4.1-m Southern Astrophysical Research (SOAR) telescope \citep{tokovinin2018ten} at Cerro Pach\'on NoirLab facility and the Very Large Telescope (VLT) at the ESO Paranal Observatory, Chile. SOAR observed \Nstar~on May 18, 2019, with the HRCam camera using the I filter (879 nm) as shown in Fig. \ref{fig:soarSpeckle}. Additionally, a high-resolution image was obtained with the NaCO instrument on the very large telescope (VLT) with adaptive optics (AO) utilizing the Ks (2.2 $\mu$m) filter on June 29, 2019, and shown in Fig. \ref{fig:NaCoAO}. The differential contrast was estimated to be $\sim$5.4 mag and  $\sim$5 mag at 1 arcsec for HRCam/SOAR and NaCO/VLT-UT4, respectively, thus providing evidence that \Nstar~is likely single.
The high-resolution speckle imaging combined with transits from TESS, LCOGT-SAAO, and NGTS missions and RVs from HARPS and FEROS provides confidence that \Nplanet\, signal has planetary origin.

%

\section{Data Analysis}
\label{sec:analysis}
\subsection{Photometric Vetting with \texttt{DAVE}}
\label{sub:vetting}
We employed the \texttt{DAVE} pipeline (Discovery and Vetting of Exoplanets, \citealt{Kostov2019}) to carefully analyse transit events on two layers of scrutiny: by looking at the image pixels and by studying the light curve.
The \texttt{centroid} module generates an image by subtracting the image taken during the transit from the one taken outside the transit. It then fits a model to this difference image to find where the light is coming from. This helps us to check whether the dip in brightness really originates from the target star or from a nearby star. If the photocenter shifts away from the target star’s position, it could mean the signal is a False Positive (FP). However, we only see a clear offset if the shift stays within the aperture mask used to retrieve the transit.
The \texttt{Modelshift} module creates a phase-folded lightcurve using a trapezoid-shaped model of the transit. Its main goal is to check whether the signal might come from an eclipsing binary. It allows to inspect features like the shape of the transit or whether odd and even dips are different, which would hint at a binary system.
\texttt{DAVE} also generates some extra diagnostics to help when the target is particularly tricky. The Lomb-Scargle Periodogram looks for repeating patterns in the light curve, which might come from a star’s rotation or from nearby stars. These effects can be caused by things like ellipsoidal variation in binary systems \citep{Morris1993, Faigler2011, Shporer2017}. When strong variations are present, we first remove them (detrend the light curve) to get a cleaner analysis.

\begin{figure}
    \centering
    \includegraphics[width=\columnwidth]{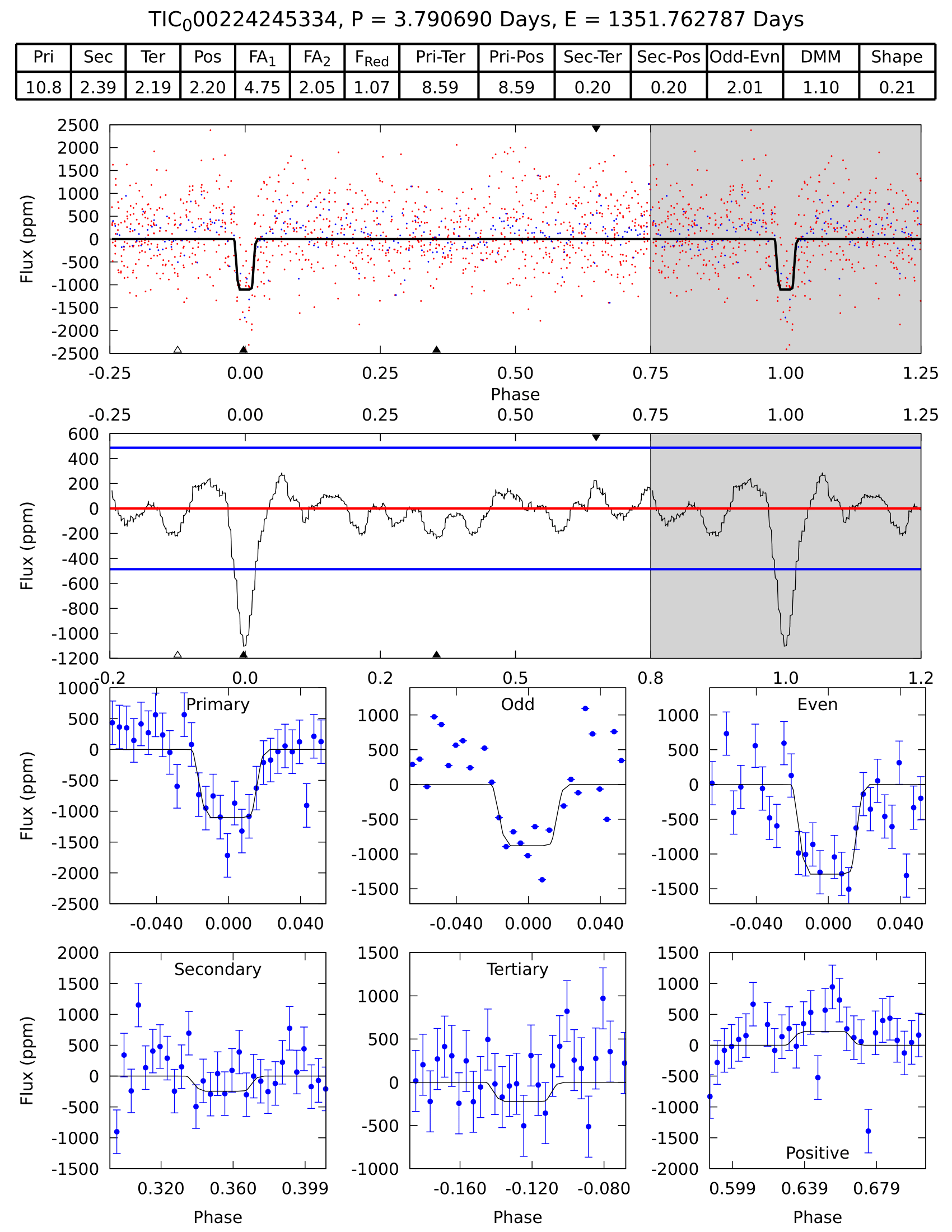}
    \caption{\texttt{Modelshift} module of TOI 333b obtained in TESS sector 02. The first panel displays the phase-folded lightcurvewith the best-fit trapezoid transit model (black line). The second panel shows the convolved lightcurvewith the transit model and noise level (blue lines). The lower panels offer close-ups of primary and secondary events, both odd and even primaries, and any additional events. The upper table indicates the statistical significance of these features, highlighted in red when flagged as significant by the pipeline.}
    \label{fig:333_modshift}
\end{figure}

\begin{figure}
    \centering
    \includegraphics[width=\columnwidth]{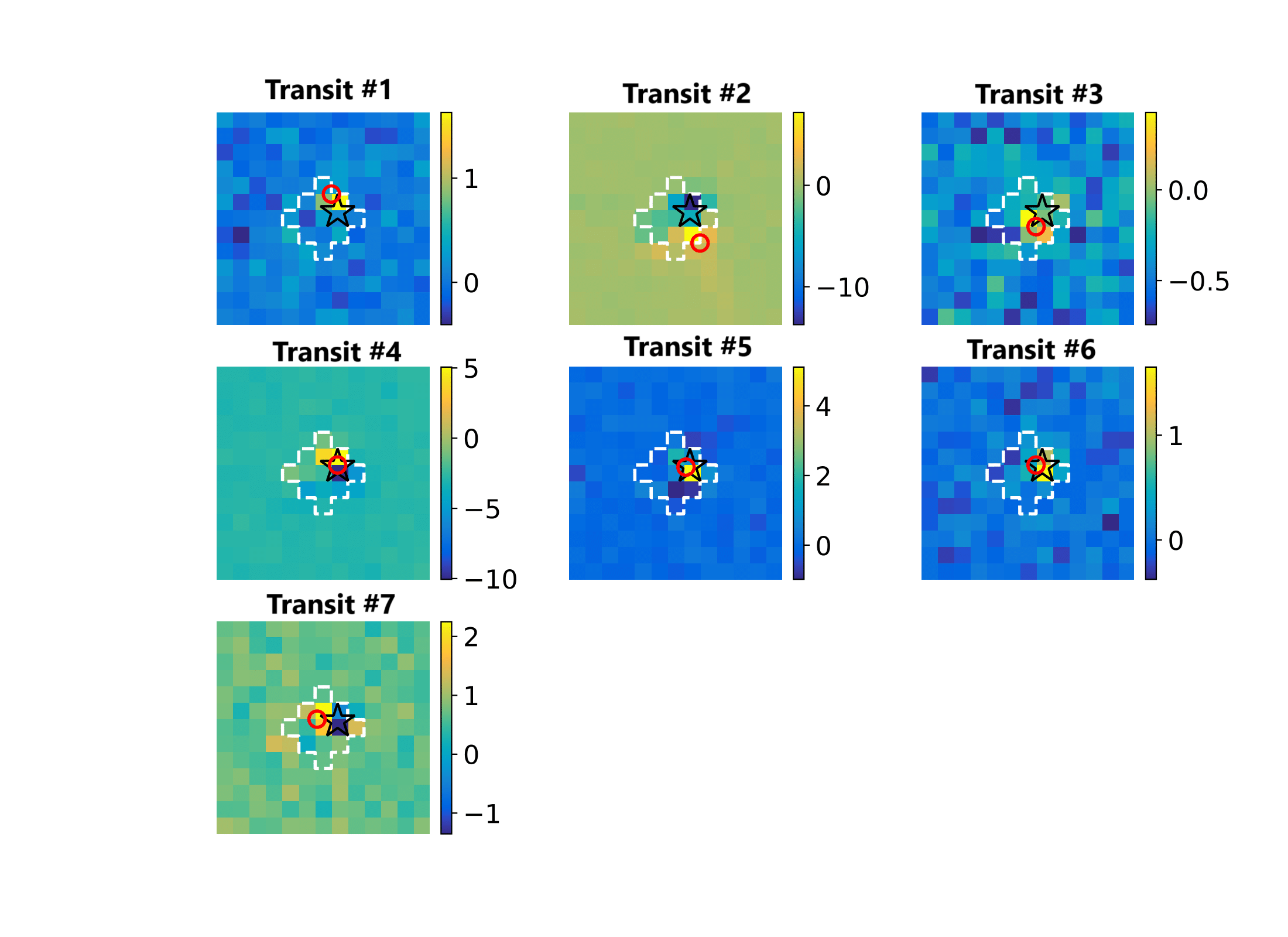}
    \caption{\texttt{centroids} module for the planet candidate TOI 333b for the 7 transits detected in TESS sector 02. The dashed white lines outline the aperture mask for lightcurve extraction. A star symbol indicates the catalogued position of the target, while individual photocentres are marked by small red dots. A colour bar indicates the number of electrons/sec for each case mentioned. A centroid image is reliable whenever no artifacts occur within the aperture mask like for transits $1,3,5$ and $6$.}
    \label{fig:333_centroids}
\end{figure}

We vetted the transit events of TOI 333b occurred in the TESS sectors $02, 29$ and $69$ by using \texttt{DAVE}. It passed all the tests as a bonafide planet candidate. We report the \texttt{Modelshift} and the \texttt{centroids} modules for TOI 333b obtained for TESS sector 02, respectively, in Fig. \ref{fig:333_modshift} and \ref{fig:333_centroids}.
The \texttt{Modelshift} returns a clear primary transit well above the noise level without any statistical significant difference between odd and even transits. Moreover, there is no evidence of a secondary feature that could have been caused by the occultation had \Nplanet\,been an eclipsing binary. As far as the \texttt{centroids} module is concerned, the overall photocenter is consistent with the position of the target despite a negligible offset due to contamination of nearby resolved sources. However, the brightest pixel in the difference image perfectly corresponds with the target's position. 
The \texttt{Lomb-Scargle Periodogram} did not highlight any modulation of the lightcurvecompatible with ellipsoidal variations typical of a binary star system.

Finally, we note that TESS has a large pixel size, about $21''$ per pixel, and a slightly blurred focus, which means light from nearby or background stars can mix into the same pixel. This can cause problems even when the \texttt{centroid} module shows no shift in the image. Other stars within a single pixel can still affect the transit signal. In some cases, this extra light can make the transit look shallower than it really is, leading to an underestimated planet size. In worse cases, the transit might not come from the target star at all.
To avoid these issues, we checked star catalogues \citep{Wenger2000, GaiaCollaboration2021} to look for any nearby stars within the aperture used to extract the light curve. We found no unresolved stars that could contaminate the light from TOI 333, confirming the transit signal is clean.
%
%
\subsection{Stellar Properties}
\label{sub:stellar}
TOI-333 stellar parameters and chemical abundances were computed with the spectroscopic parameters and atmospheric chemistries of stars \citep[\texttt{SPECIES}\footnote{github.com/msotov/SPECIES};][]{soto2018spectroscopic} and the spectral energy distribution Bayesian model averaging fitter \citep[\texttt{ARIADNE\footnote{https://github.com/jvines/astroARIADNE}};][]{vines2022ariadne} from spectroscopic analysis and archival photometry spectral energy distribution (SED) fitting, respectively. Finally, we have independently derived stellar parameters with PHOENIX stellar atmosphere models and EXOFASTV2.
\subsubsection{\texttt{SPECIES}+\texttt{ARIADNE}}
\label{subsub:species+ariadne}
Atmospheric parameters such as effective temperature (T$_{\rm eff}$), metallicity [Fe/H], surface gravity ($\log g$), and micro-turbulence velocity ($\xi_t$) were estimated from HARPS high resolution spectra using the equivalent width (EW) method implemented in \texttt{SPECIES}. The coadded spectra is handed to \texttt{SPECIES}, which computed the EW for Fe I and Fe II lines. Astrometric and photometric archival data are fetched from databases like Vizier\footnote{https://vizier.cds.unistra.fr/viz-bin/VizieR} and used to compute the initial guesses for the atmospheric parameters. The starting values, EW and a grid of atmospheric models from ATLAS9 \citep{castelli2004new} are introduced to MOOG \citep{sneden1973nitrogen}, which solves the radiative transfer equation (RTE) while measuring the correlation between Fe line abundances as a function of excitation potential and EW, assuming local thermodynamic equilibrium (LTE). During the search for the RTE solution, an iterative process is carried out up to 10,000$\times$ or until atmospheric parameters that lead to no correlation between the iron abundances and the excitation potential and the reduced equivalent width (EW$/\lambda$), are found. The adopted T$_{\rm eff}$, [Fe/H],  $\log g$ along with parallax, photometry in several bands, and proper motions are used to compute the mass, radius and age from the \texttt{isochrone} package \citep{morton2015isochrones} by interpolating through a grid of MIST \citep{dotter2016mesa} evolutionary tracks. Nested sampling \citep[][]{feroz2009multinest} is used to properly estimate posterior distributions for \mstar,\,\rstar, and age. Finally, \texttt{SPECIES} estimates stellar rotation ($v\sin{i}$) and macro turbulent velocities from temperature calibrators and fitting the absorption lines of observed spectra with synthetic line profiles. From our analysis using \texttt{SPECIES} we derived the following stellar properties for \Nstar\,with median and 1-$\sigma$ confidence intervals: T$_{\rm eff} = 6267 \pm 50$\,K, $\text{[Fe/H]} = 0.00 \pm 0.05$\,dex, $\log \text{g} = 4.42 \pm 0.08$, and $v\sin{i} = 6.18 \pm 0.70$\, km/s. Chemical abundances are given in Table \ref{tab:CheAb}. 

We used the publicly available \texttt{ARIADNE} \citep{vines2022ariadne} python tool to derive parameters for \Nstar. The code is based on the SED fitting method, which consists of fitting archival photometry to synthetic magnitudes from interpolated grids of stellar atmosphere models. The synthetic photometry is computed by convolving a given model with several filter response functions \citep[see available SED models in][]{vines2022ariadne} and scaled by $(R/D)^{2}$. An excess noise term is introduced for each photometric measurement in order to account for underestimated uncertainties. Finally, a cost function with input parameters T$_{\rm eff}$, \logg, \met, and $V$ band extinction (A$_{\rm V}$) is explored with \citep[\texttt{DYNESTY}][]{speagle2020dynesty}, a nested sampling algorithm used to effectively search the parameter space global minimum, thus finding the best set of synthetic fluxes from a given SED model with stellar properties that best matches the observed photometry. \\
\texttt{ARIADNE} performs the steps laid out above for several atmosphere libraries, from which we used Phoenix V2 \citep{husser2013new}, BT-Settl \citep{hauschildt1999nextgen,allard2012models}, \citet{castelli2004new}, and Kurucz \citet{kurucz1993atlas9} models. The estimated stellar posterior distributions from each library are weighted by their Bayesian evidences and averaged to build the adopted final stellar parameters distributions. The Bayesian averaging method helps mitigate biases and uncertainties coming from the assumptions and limitations from individual stellar atmosphere models, thus yielding precise stellar parameters, particularly the \rstar\, and T$_{\rm eff}$, which are key to inform the global modelling of \Nplanet\ (see $\S$ \ref{sub:globalmodelling}). Finally, T$_{\rm eff}$, \logg, \met\,as well as additional quantities such as D, \rstar, and A$_{\rm V}$ from \texttt{ARIADNE} are used to derive the stellar age (\age), \mstar, and the equal evolutionary points from the \texttt{isochrone} package.

We started \texttt{ARIADNE} with priors defined in Table \ref{tab:priors-ariadne}, centred on the \texttt{SPECIES} posterior distributions but with slightly broader variances to allow a more extensive exploration of the parameter space. $\mathcal{N}$($\mu$,$\sigma^2$) represents a Gaussian prior with $\mu$ and $\sigma$ as mean and variance, respectively. \texttt{ARIADNE} best-fitting SED model is shown in Fig. \ref{fig:sed}, with adopted stellar parameters median and 1$-\sigma$ in Table \ref{tab:stellar} along with the archival photometry. The SED shows two observed magnitudes (cyan) that deviate from the expected model (black) and the synthetic magnitudes (purple) around 0.8–0.9~$\mu$m. We reran \texttt{ARIADNE} excluding these photometric bands and confirmed that their removal does not yield statistically significant changes in the derived stellar parameters. Therefore, we retained all available photometric data.

We compared our results with GAIA DR3\footnote{https://gea.esac.esa.int/archive/}\citep{vallenari2023gaia} stellar parameters, which are in agreement with our adopted values from \texttt{ARIADNE}. GAIA parameters are as follows, T$_{\rm eff}$ = 5990 $^{+5}_{-2}$\,K, log g = 4.341$\pm 0.004$\,dex, \rstar\ = 1.15$\pm 0.01$\,\rsun\,, $\age = 4.9 \pm 0.8$\,Gyr , distance of 347$\pm 2$\,pc, and A$_{\rm V}$ = 0.002$^{+0.003}_{-0.001}$. Finally, we adopted the \texttt{ARIADNE} stellar parameters for \Nstar, given its robust Bayesian framework that combines multiple atmospheric model grids with nested sampling, efficiently exploring the parameter space and mitigating limitations inherent to individual atmospheric libraries. The derived results are consistent with the independent analysis presented in $\S$~\ref{subsubsec:IndAnalysis}.
\subsubsection{Independent Stellar parameters analysis}
\label{subsubsec:IndAnalysis}
\begin{itemize}
    \item Stellar Parameters with PHOENIX Models
\end{itemize}
We performed an analysis of the broadband SED of the star together with the {\it Gaia\/} DR3 parallax \citep[with no systematic offset applied; see, e.g.,][]{StassunTorres:2021}, in order to determine an empirical measurement of the stellar radius, following the procedures described in \citet{Stassun:2016,Stassun:2017,Stassun:2018}. We pulled the $JHK_S$ magnitudes from {\it 2MASS}, the $G G_{\rm BP} G_{\rm RP}$ magnitudes from {\it Gaia}, the $B_T V_T$ magnitudes from {\it Tycho-2}, the W1--W3 magnitudes from {\it WISE}. We also utilised the NUV magnitude from {\it GALEX} and the absolute flux calibrated spectrophotometry from {\it Gaia}. Together, the available photometry spans the full stellar SED over the wavelength range 0.4--20~$\mu$m (see Figure~\ref{fig:phoenix-SED}).
 
\begin{figure}
	\includegraphics[width=\columnwidth]{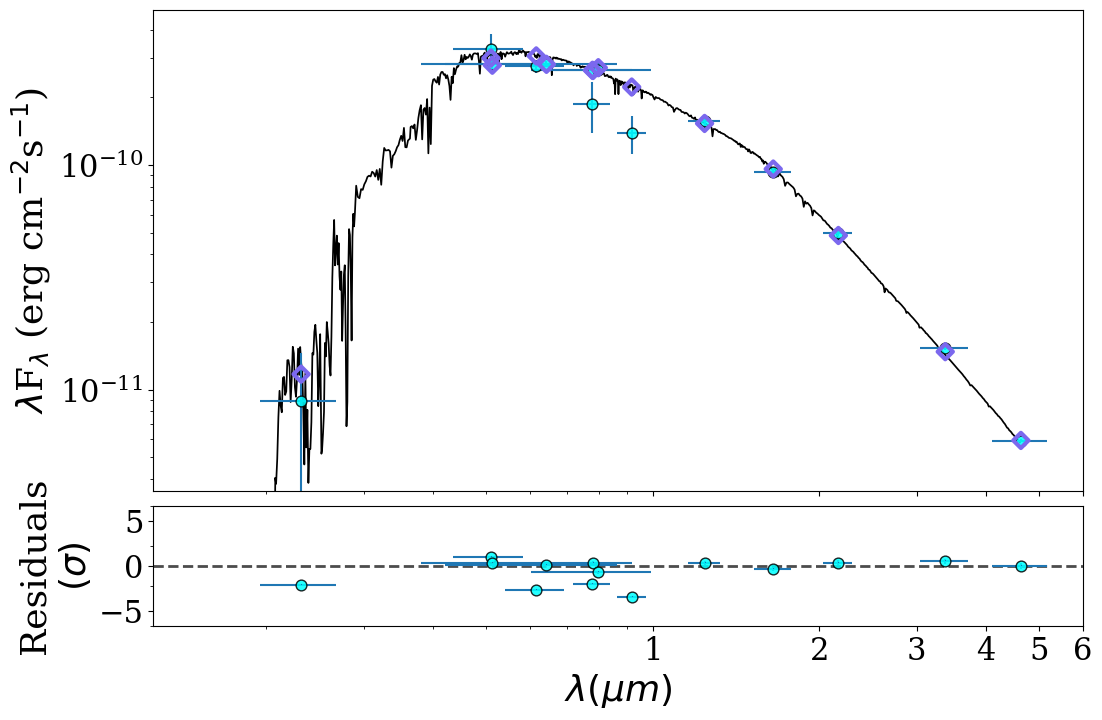}
    \caption{\textbf{Top}: The best-fitting spectral energy distribution (black line) based on \citet{castelli2004new} given the TOI-333 photometric data (cyan points) and their respective bandwidths shown as horizontal errorbars. Purple diamonds represent the synthetic magnitudes centred at the wavelengths of the photometric data from Table~\ref{tab:stellar}. \textbf{Bottom}: residuals to the best fit in $\sigma$ units.}
    \label{fig:sed}
\end{figure}

We performed a fit using PHOENIX stellar atmosphere models \citep{husser2013new}, with the $T_{\rm eff}$, $\log g$, and metallicity ([Fe/H]) adopted from the spectroscopic analysis. The extinction, $A_V$, was limited to maximum line-of-sight value from the Galactic dust maps of \citet{schlegel1998maps}. The resulting fit (Figure~\ref{fig:sed}) has a reduced $\chi^2$ of 1.9, with a best-fit $A_V = 0.04 \pm 0.02$. Integrating the (unreddened) model SED gives the bolometric flux at Earth, $F_{\rm bol} = 4.290 \pm 0.100 \times 10^{-10}$ erg~s$^{-1}$~cm$^{-2}$. Taking the $F_{\rm bol}$ together with the {\it Gaia\/} parallax directly gives the bolometric luminosity, $L_{\rm bol} = 1.69 \pm 0.05$~L$_\odot$. The stellar radius follows from the Stefan-Boltzmann relation, giving $R_\star = 1.13 \pm 0.03$~R$_\odot$. In addition, we can estimate the stellar mass from the empirical relations of \citet{torres2010accurate}, giving $M_\star = 1.17 \pm 0.07$~M$_\odot$.

Finally, we can estimate the stellar rotational velocity from the spectroscopic $v\sin i$ together with $R_\star$, giving $P_{\rm rot} / \sin i = 8.5 \pm 0.9$ days, from which we can use empirical gyrochronology relations \citep{mamajek2008improved} to estimate an age of $1.0 \pm 0.2$~Gyr.

\begin{itemize}
    \item \texttt{EXOFASTv2} analysis
\end{itemize}

As an additional constraint on the stellar parameters, we performed an analysis using the \texttt{EXOFASTv2}\footnote{\url{https://github.com/jdeast/EXOFASTv2}} modelling suite \citep{Eastman:2019}. This tool simultaneously fits the SED and the MESA Isochrones and Stellar Tracks (MIST) models \citep{Dotter:2016, Choi:2016}. The SED was constructed using available broadband photometry. We adopted the parallax from {\it Gaia} DR3, corrected for the global offset following \citet{Lindegren2021}. An upper limit on the $V$-band extinction, $A_V$, was imposed using the dust maps of \citet{Schlafly:2011}.
Gaussian priors were placed on $T_{\rm eff}$, log $g$, and [Fe/H] based on the spectroscopic analysis described in the subsection~\ref{subsub:species+ariadne}.
The resulting stellar parameters are: $T_{\mathrm{eff}} = 6205 \pm 45$, K, $R_\star = 1.09 \pm 0.02,R_\odot$, and $M_\star = 1.05 \pm 0.05 M_\odot$.

As an additional check, we repeated the EXOFASTv2 analysis replacing the MIST stellar models with the PARSEC models \citep{Bressan2012}, keeping all other inputs and priors identical. The resulting stellar parameters from the SED+PARSEC fit are fully consistent with the MIST-based values, with
$T_{\mathrm{eff}} = 6207 \pm 44$, K, $R_\star = 1.09 \pm 0.02~R_\odot$, and $M_\star = 1.07 \pm 0.05~M_\odot$.

\subsubsection{Insights into TOI-333’s Age}
\label{subsub:Agestimation}
Stellar ages are one of the key ingredients to understand how the ND forms. Proposed scenarios for their formation and evolution histories frequently include a combination of planet migration, envelope mass loss through photoevaporation, tidal disruption or RLO. The physical processes take place at distinct timescales; thus, a reliable age estimate, along with several other planetary system parameters, such as mass, radius, and bulk densities, particularly their ages provide insights into how ND planets form and evolve. 

A typical method used to derive stellar ages consists of using grids of pre-computed stellar evolutionary models described by stellar physical properties such as \teff, \lstar, and \met\,that are interpolated to fit a set of observables. Such evolutionary models are then rearranged to tracks of fixed ages named isochrones, from which stellar ages are estimated. However, the complexity and strong non-linearity of isochrones, along with observational uncertainties, make it difficult to precisely estimate stellar ages for stars on or near the main sequence.

\texttt{ARIADNE}'s age of 0.57$^{+3.72}_{-0.54}$ Gyr is computed using the \texttt{isochrone} package, which to 1-$\sigma$ indicates a broad age range from 30 Myr to 4.29 Gyr. In order to increase precision, we assessed \Nstar's age based on other methods such as (1) the rotation-age correlation, also known as Gyrochronology, which assumes that stellar ages are, to first order, correlated with the rotation period, and (2) the Lithium abundance-age correlations, while owing to lithium volatility with temperature, its abundance is quickly depleted in stellar atmospheres within a few hundred million years of a star’s lifetime depending on its initial mass \citep[e.g.,][]{christensen2018ages}.

For the Gyrochronology method, we compared the models by \citet{barnes2007ages, mamajek2008improved}, and \citet{meibom2009stellar} to the P$_{\rm rot}/\sin{i}$ derived from $v\sin{i}$ and \rstar\, from Table \ref{tab:stellar} as a function of colour B-V. The models point towards \Nstar\,having gyro-ages in the range of 0.7 to 2.2 Gyrs depending on the selected model, thus in statistical agreement within 1-$\sigma$ from the adopted age from \texttt{ARIADNE}, respectively. Moreover, we searched \texttt{TESS} photometry for brightness variation caused by spots appearing on the stellar disk due to the star's spin. We first masked the transits, median normalised, and binned the time-series to 30 minutes. We then employed the Lomb-Scargle \citep[LS;][]{vanderplas2018understanding} as well as the Edelson-Krolik Auto-Correlation Function \citep[ACF;][]{vanderplas2012introduction} methods independently in order to search for a photometric rotation period for the joint TESS sectors. The LS 1$^{\rm st}$ and 2$^{\rm st}$ highest peaks were $\sim 4.98$ at normalised power $\sim 0.02$ and $\sim 9.43$ days at $\sim 0.014$, respectively, yet no significant variability was found in the folded photometry at those peaks within $\sim 120$ ppm photometric precision, and a false alarm probability at the highest peak of $\sim 6 \times 10^{-16}$. The periodograms were also examined on a per-sector basis, but no significant peaks were detected. Finally, although the lightcurves did not present oscillatory patterns, mild variability had been visually detected (see Fig. \ref{fig:all-lc} upper panel), possibly caused by high latitude spots.

We identified the presence of lithium in \Nstar\,spectra at 6707.856 \AA. We measured the Li line EW from fitting a double Gaussian model to the Li and the nearby partially blended FeI 6707.4 \AA~lines. Fig. \ref{fig:Lithium} shows the normalised coadded HARPS spectrum around the Li $\lambda$ 6707.856 \AA\,line. We obtained a value of EW = $85.54^{+1.15}_{-1.26}$m\AA\, from the Li-only Gaussian model in red, where the 1-$\sigma$errors were derived from bootstrapping with 50,000 iterations. The presence of Lithium provides an independent means to provide an upper limit on the system's age. Finally, in Fig.~\ref{fig:ew-teff}, we compare the \Nstar\,EW measurement to stars of the same spectral type in open clusters, where age estimates are generally more reliable over field stars. The data is sourced from \citet[][]{albarran2020gaia}, which we have adapted here to display a few clusters, though a comparison can also be made with Fig. 10 in their study. While a detailed Li-age analysis is beyond the scope of this work, given that lithium depletion is influenced by factors including stellar rotation and metallicity, Fig.~\ref{fig:ew-teff} suggests that the majority of late F-type stars in older clusters ($> 1$ Gyr) typically exhibit greater lithium depletion compared to \Nstar\, allowing us to place an upper age limit of approximately $\sim 1$ Gyr. This estimate is consistent with the results from \texttt{ARIADNE} and aligns with expectations from Gyrochronology models. These indicators suggest that \Nstar\, is likely young, thus we adopted an age upper limit of $\sim 1$ Gyr for the planetary system.
%
\begin{table}
	\centering
	\caption{Stellar Properties for TOI-333}
	\begin{tabular}{lcc} 
	Property	&	Value		&Source\\
	\hline
    \multicolumn{3}{l}{Astrometric Properties}\\
    R.A.		&	\mbox{$20^{h} 45^{m} 01\fs9941$}			&GAIA\\
	Dec			&	\mbox{$-35\degr 25\arcmin 40\farcs 2322$}			& GAIA	\\
	2MASS I.D.	& J23332579-4110174	&2MASS	\\
	TIC I.D.	& 224245334	&TIC	\\
	GAIA DR3 I.D. & 6535747220036134272	&GAIA	\\
	Parallax (mas) & 2.825 $\pm$ 0.017 &GAIA \\
    $\mu_{{\rm R.A.}}$ (\masy) & 16.607 $\pm$ 0.017 & GAIA \\
	$\mu_{{\rm Dec.}}$ (\masy) & 1.562 $\pm$ 0.016 & GAIA \\
    \\
    \multicolumn{3}{l}{Photometric Properties}\\
	V (mag)		&11.991 $\pm$ 0.013 &APASS\\
	B (mag)		&12.551 $\pm$ 0.023	&APASS\\
	g (mag)		&12.218	$\pm$ 0.009	&APASS\\
	r (mag)		&11.888 $\pm$ 0.010	&APASS\\
	i (mag)		&11.779 $\pm$ 0.026	&APASS\\
    G (mag)		&11.8993 $\pm$ 0.0003		& GAIA\\
    TESS (mag)	&11.517 $\pm$ 0.006	&TIC\\
    J (mag)		&10.982 $\pm$ 0.026 		&2MASS	\\
   	H (mag)		&10.759 $\pm$ 0.023 	&2MASS	\\
	K (mag)		&10.678 $\pm$ 0.021	&2MASS	\\
    W1 (mag)	&10.645 $\pm$ 0.022 	&WISE	\\
    W2 (mag)	&10.697 $\pm$ 0.02 	&WISE	\\
    W3 (mag)	&10.696 $\pm$ 0.094	&WISE	\\
    
    \\
    \multicolumn{3}{l}{Derived Properties}\\
    $\rho_{*}$ (\gccc)   & 1.05$^{+0.05}_{-0.06}$   &\texttt{Juliet}\\
    $\gamma_{RV-HARPS}$ (\kms) & 0.296$\pm$0.002&\texttt{Juliet}\\
    $\gamma_{RV-FEROS}$ (\kms) & 0.279$\pm$ 0.005&\texttt{Juliet}\\
    P$_{\rm rot}/\sin{i}$ (days) &  9.01$^{+1.43}_{-1.14}$ & This work\\
    $v\sin{i}$ (\kms) &  6.18 $\pm$ 0.70 & \texttt{SPECIES}\\
    T$_{\rm eff}$ (K)    & 6241	$^{+73}_{-62}$    &\texttt{ARIADNE}\\
    $[$Fe/H$]$	 & 0.01	$\pm$ 0.04       &\texttt{ARIADNE}\\
    log g                &	4.41 $\pm$ 0.08  &\texttt{ARIADNE}\\
    Age	(Gyr)		     & < 1 Gyr  & This work \\
    \mstar (\msun)       & 1.2 $\pm$ 0.1 	 &\texttt{ARIADNE}\\
    \rstar (\rsun)       & 1.10 $\pm$ 0.03 	      &\texttt{ARIADNE}\\
    Distance (pc)	     & 347 $\pm$ 8   &\texttt{ARIADNE} \\
	\hline
    \multicolumn{3}{l}{2MASS \citep{2MASS}; TIC v8 \citep{stassun2018tess};}\\
    \multicolumn{3}{l}{APASS \citep{APASS}; WISE \citep{WISE};}\\
    \multicolumn{3}{l}{{\em Gaia} \citep{brown2021gaia}}\\
	\end{tabular}
    \label{tab:stellar}
\end{table}

\subsection{Global modelling}
\label{sub:globalmodelling}
A joint radial-velocity and photometric analysis has been performed with the \texttt{Juliet} \citep{espinoza2019juliet} package, a versatile code wrapped around packages like \texttt{Batman} \citep{kreidberg2015batman} for lightcurve modelling and \texttt{radvel} \citep{fulton2018radvel} for RV analysis. The dataset is comprised of 37 HARPS, 7 FEROS RVs, and a total of 32,282 photometric data points from TESS, LCOGT-SAAO and NGTS.

Each instrument has its own precision and operates under different environmental conditions, leading to variations in how noise is encapsulated in each dataset. As a result, proper modelling is essential to optimally derive planetary properties. To address this, we incorporated Gaussian Processes (GP) into the noise model to account for correlated noise in the lightcurves, modelling each instrument with an approximate Mat\'ern kernel. While a global GP kernel is generally preferred for accurately modelling stellar activities, we opted for a multi-instrument GP approach because \Nstar\,exhibits moderately low activity in the photometry and RVs highlighted in Fig. \ref{fig:all-lc}. No GP was applied to the RVs as shown by the Keplerian component of the global model in the bottom panel. We note that the GLS did not present any significant peak outside the planet period, which could be potentially attributed to activities in the RVs.

No dilution term was introduced as the transit depths from different missions are in statistical agreement, as highlighted in Fig. \ref{fig:ttvs-tdvs-dv} lower panel or Table \ref{tab:ttvs}. We note that the transit durations are also consistent amongst distinct instruments. Thus, we conclude that no dilution treatment is necessary. For limb darkening, we followed the approach outlined in \citet{kipping2013efficient}, employing a quadratic parameterisation with $q_1$ and $q_2$, where uniform priors $\mathcal{U}(0, 1)$ were assigned for each instrument. 

The radial-velocity part of the global model includes a Keplerian, a systemic RV term ($\gamma_{RV}$) and a white noise term to account for stellar jitter. The eccentricity $e$ and the argument of periapsis EW were fixed to zero based on model comparison metrics like their evidence factor ln(Z). The model with fixed $e$ and $\omega$ is favoured by a log-evidence difference of $\Delta\ln Z = 6$ compared to the model where $e$ and $\omega$ are allowed to vary under uniform priors $\mathcal{U}(0.0, 0.1)$ and $\mathcal{U}(0.0, 90^{\circ})$, respectively.
Nonetheless, we note that computing \Nplanet\,eccentricity given our dataset is challenging due to a combination of the systems' small planetary signal amplitude and the mean RV precision. Therefore, our run with free eccentricity provides an upper limit of 0.03 at 1-$\sigma$, thus further evidencing \Nplanet\,likely circular orbit. Finally, due to the high dimensionality of the parameter space, we used the dynamic nested sampling algorithm \citep{higson2019dynamic} through \texttt{DYNESTY} with 1,300 live points to explore the parameter space, where we derived the solution shown in Table \ref{tab:planet}.

\begin{table}
	\centering
	\caption{Planetary Properties for \Nplanet}
	\begin{tabular}{lc} 
	Property	&	Value \\
	\hline
    P (days)		        & 3.7852503 $\pm$ 0.0000057	\\
	T$_C$ (BJD$_{\rm TDB}$)		& 2458355.5759 $\pm$ 0.0024	 \\
    T$_{14}$ (hours) & 3.20 $\pm$ 0.05 \\
    $a/\rstar$		        & 9.26 $\pm$ 0.18            \\
    \rpl/\rstar & 0.035 $\pm$ 0.001 \\
    $b$ & 0.12 $\pm$ 0.11                       \\
    $i(deg)$ & 89.26 $\pm$ 0.67                      \\

	K (\ms) 	&7.90 $\pm$ 0.87	                           \\
    e 			& 0.0 (fixed)  	\\
    $\omega~(\deg)$ & 90 (fixed) \\
   Jitter (\ms)& 1.59$^{+4.15}_{-1.57}$                     \\
    \mpl (\mearth)& 20.1 $\pm$2.4	\\
    \rpl (\rearth)& 4.26$\pm$0.11  \\
    $\rho_{p}$ (\gccc) & 1.42$\pm$0.21 \\
    a (AU) & 0.049$\pm$0.001 \\
    T$^{\ast}_{eq}$ (K) & 1445 $\pm$ 18
	\\
	\hline
    \multicolumn{2}{l}{$\ast$ Assumed zero Bond albedo}\\
	\end{tabular}
    \label{tab:planet}
\end{table}
\subsection{Transit Timing, Duration, and Depth Variation Analysis}
\label{sub:ttv-tdv-modelling}
Transit timing variations \citep*[TTVs;][]{agol2005detecting} arise when observed mid-transit times $T_0$ depart from the predicted values of a linear ephemeris, expressed as $T_n = T_0 + N \times P$, where $N$ represents the transit number and $P$ the orbital period. In other words, the difference between the observed and computed transit times defines the TTV points; a statistically significant deviation from zero would indicate that the planet's orbit has been altered. Two primary causes for TTVs are commonly identified in the literature: (1) dynamical interactions between the planet and its host star, often resulting in angular momentum loss and orbital decay, potentially leading to the planet spiralling inward toward the star \citep[e.g., WASP-12;][]{wong2022tess}; and (2) gravitational perturbations between planets, particularly near mean motion resonance (MMR) regions, where the mutual interactions are stronger, producing larger TTV signals \citep[e.g., WASP-47;][]{becker2015wasp}. These systems frequently exhibit anti-correlated TTVs and transit duration variations (TDVs). Notably, numerous multi-planet systems near MMRs detected by Kepler were confirmed through the TTV method alone \citep{cochran2011kepler,gillon2017seven,steffen2012transit}, as their faint host stars hindered high-precision radial velocity (RV) follow-up. Nonetheless, in some cases, RVs have provided mass estimates \citep[][]{barros2014sophie, almenara2018sophie}, and in others, precise TTV measurements have enabled mass and eccentricity determinations via dynamical modelling \citep{lithwick2012extracting}, uncovering the so-called chopping effect that helps resolve the mass–eccentricity degeneracy.

\begin{figure}
	\includegraphics[width=\columnwidth]{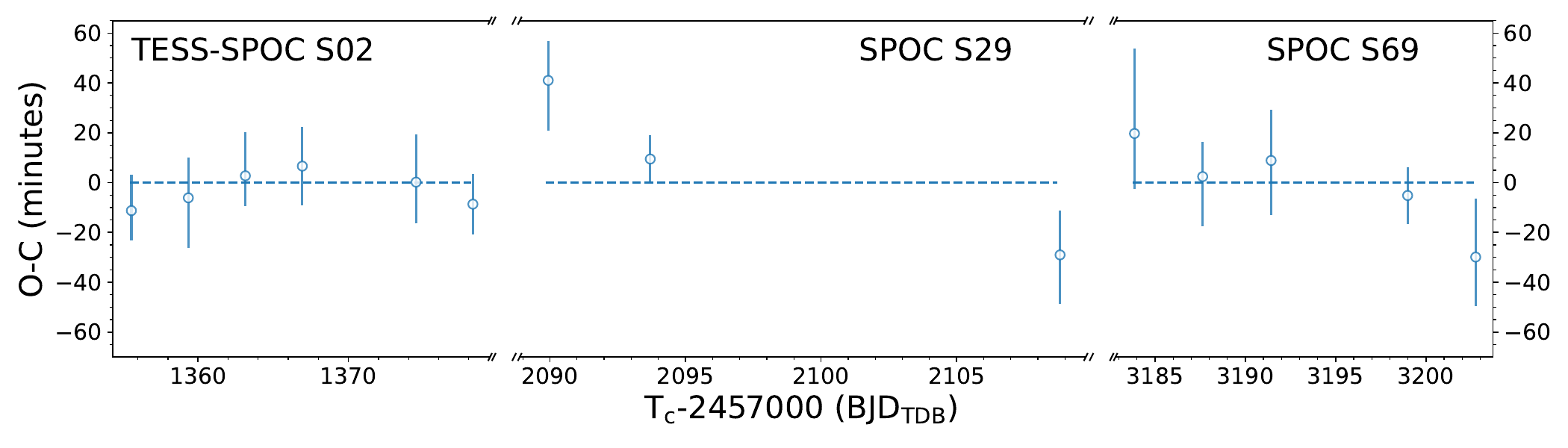}
    \includegraphics[width=\columnwidth]{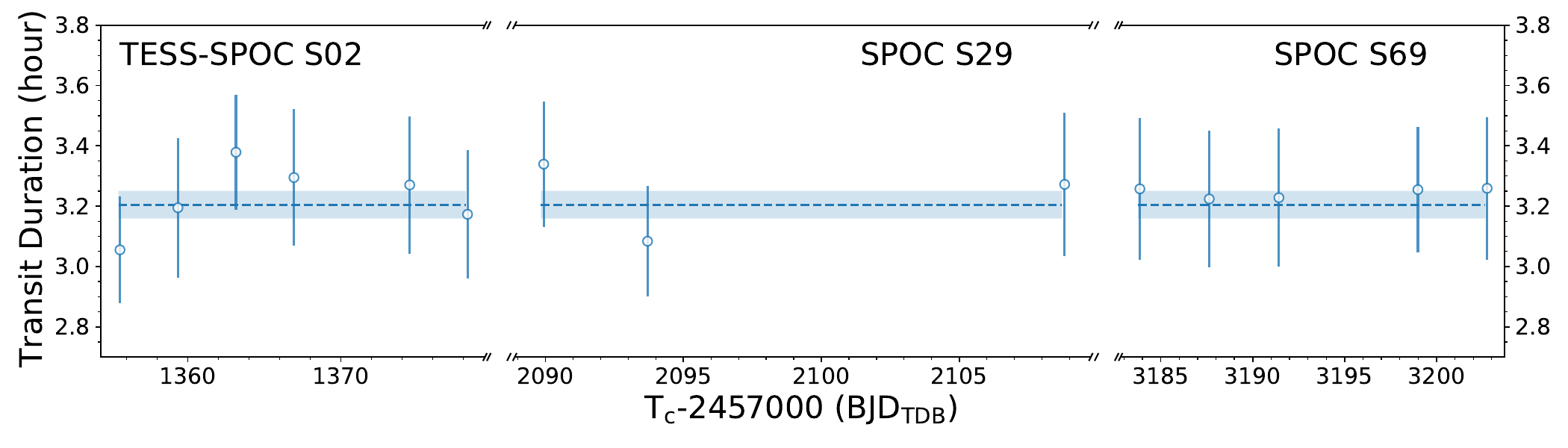}
    \includegraphics[width=1.02\columnwidth]{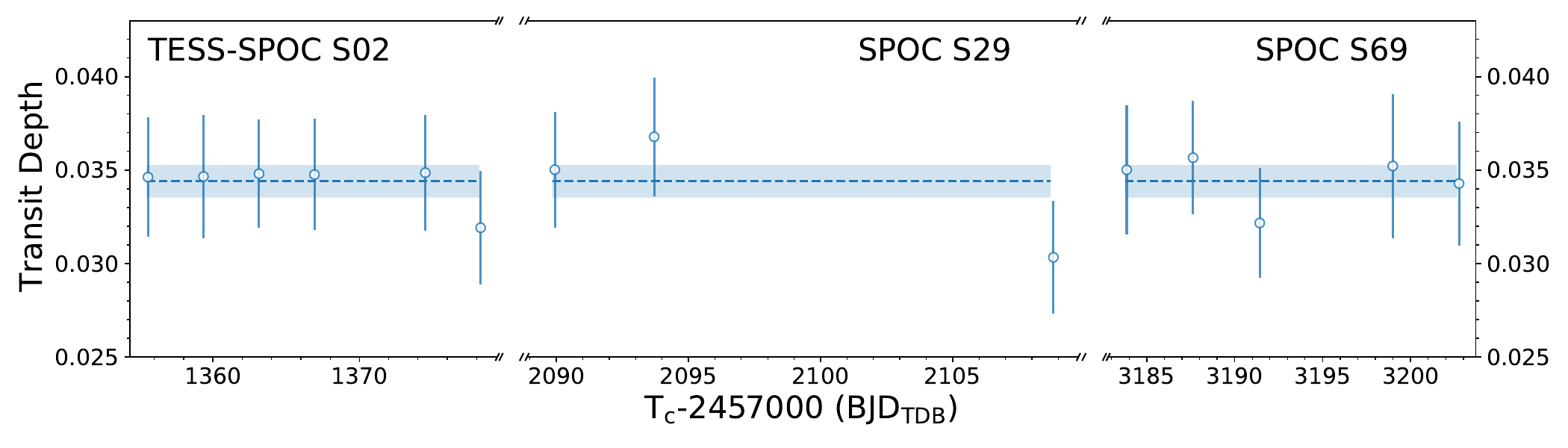}
    \caption{{\bf Top}: Transit timing variation for the TESS mission. Blue open circles represent each transit time subtracted from the best-fitting linear ephemeris. The pipelines and sectors are labelled on the top. The abscissa was zoomed for better visualisation while avoiding the large gaps in the time domain. {\bf Centre}: Transit duration variation. The dotted blue and light blue shaded region represent the transit duration median and its 1-$\sigma$ confidence interval. {\bf Bottom}: Transit depth variation. Colour scheme is the same as above.}
    \label{fig:ttvs-tdvs-dv}
\end{figure}

\Nplanet\,TTVs were estimated from the GP detrended TESS time-series using only transits with full coverage. Each transit mid-time $T_n$ was modelled with the \texttt{batman} \citep{kreidberg2015batman} code and final distributions computed with the affine invariant Markov chain Monte Carlo (MCMC) code implemented in the \texttt{emcee} package \citep{foremanmackey13}. The transit depth $p = \frac{R_b}{R_*}$, normalised semi-major axis $a = \frac{a}{R_*}$ and a linear offset $bl$ around the normalised flux were free parameters in the model. $T_n$, $p$, and $a$ were assigned the following uniform priors $\mathcal{U}$($T_n$-0.05, $T_n$+0.05), $\mathcal{U}$($p$-0.03, $p$+0.03), $\mathcal{U}$($a$-2, $a$+2), and $\mathcal{U}$($bl$-10$^{-4}$, $bl$+10$^{-4}$), respectively, with all other parameters being fixed to their medians from Table \ref{tab:planet}. Finally, a linear ephemeris model was fit to the $T_n$, where we used 10,000 MCMC steps, where 20$\%$ were discarded as burn-in. The best-fitting linear model parameters $T_0$ and $P$ for TESS are given by $T_0 = 2458355.572 \pm  0.004$ days\footnote{the unit is in barycentric Julian date in the barycentric dynamical time.} and $P = 3.785251 \pm 0.000013$ days. Using the TESS best-fitting linear ephemerides, we calculated the expected T$_n$ for NGTS, and modelled the transits observed T$_n$. The TTV points for the two NGTS transits are 24.60$^{+8.94}_{-8.33}$ min and 16.56$^{+9.19}_{-8.46}$ min at 1-$\sigma$, with the transits listed in chronological order. NGTS transit durations and depths were fixed to the best-fitting values in table \ref{tab:planet} while computing the individual transits T$_n$ since the relatively short out-of-transit data combined with photometric scatter impacted the T$_n$ computation. Likewise, LCOGT-SAAO TTVs are given by 8.28$^{+4.51}_{-4.44}$ min at 1-$\sigma$, while its transit duration and depth are 3.21$\pm$ 0.01 hours and 0.035$\pm$0.001, respectively. The transit depths, timing and duration variations are made available in Table \ref{tab:ttvs}. Fig. \ref{fig:ttvs-tdvs-dv} shows \Nplanet\, TTV points, i.e. the differences between the observed and computed transit times relative to the linear ephemeris model. No significant TTVs were detected, providing no evidence for changes in the planet’s orbit. More data and a longer temporal baseline may help further constrain the possible TTV scenarios for this planet.
\section{Discussion}
\label{sec:discussion}
\begin{figure}
	\includegraphics[width=\columnwidth]{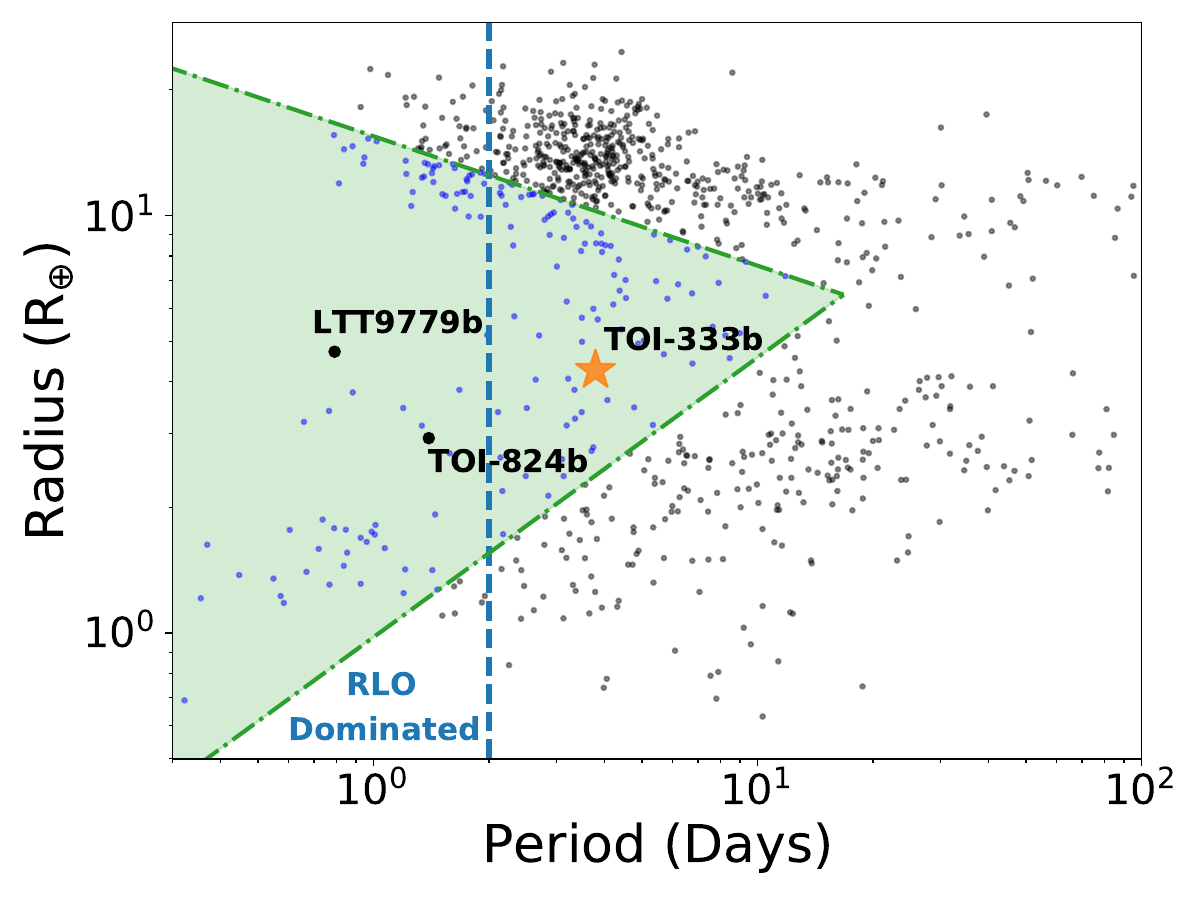}
    \caption{Radius–period diagram for well-studied exoplanets from the TEPCat catalogue \citep{southworth2011homogeneous}. \Nplanet\, is shown as a yellow star within the green triangular region defined by \citet{mazeh2016dearth}, which outlines the ND. Blue and black points represent planets inside and outside this region, respectively, with LTT9779b and TOI-824b highlighted for comparison. The dashed blue line at P $\sim$ 2 days marks the orbital period below which RLO is thought to significantly influence atmospheric evolution in the desert.}
    \label{fig:per-rad-diag}
\end{figure}
Here we place \Nplanet\,into context with the ND population, examine its possible internal structure, and assess its potential for atmospheric characterisation with the James Web Space Telescope (JWST).
\subsection{\Nplanet\,and the Neptune Desert population}

\begin{figure}
	\includegraphics[width=\columnwidth]{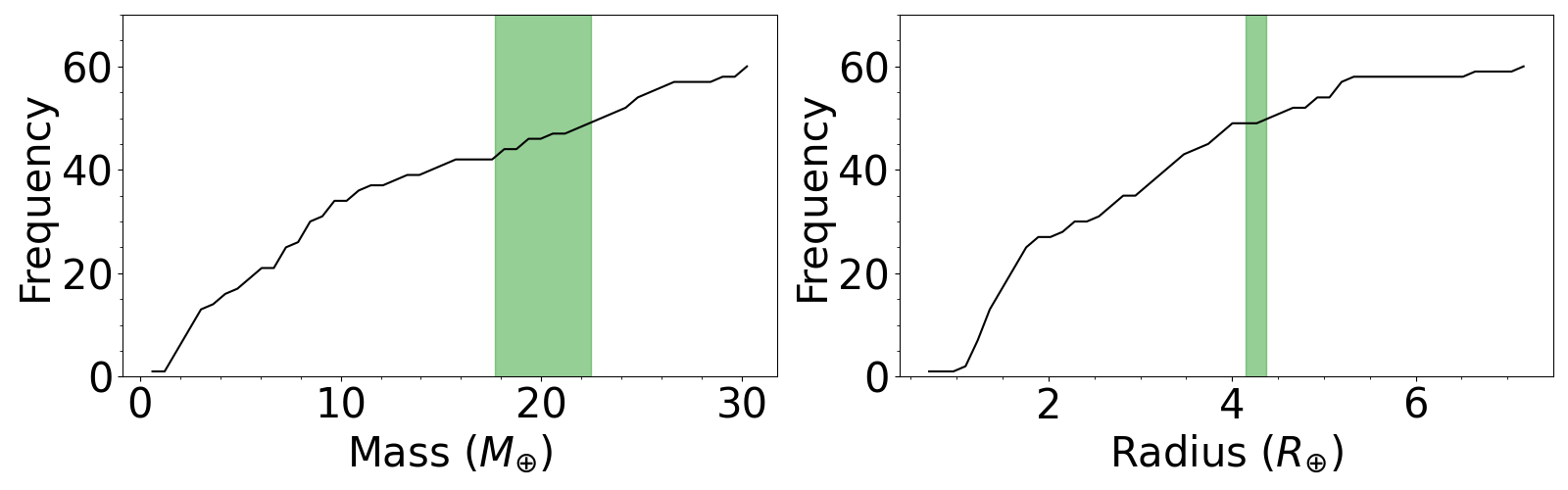}
	\includegraphics[width=\columnwidth]{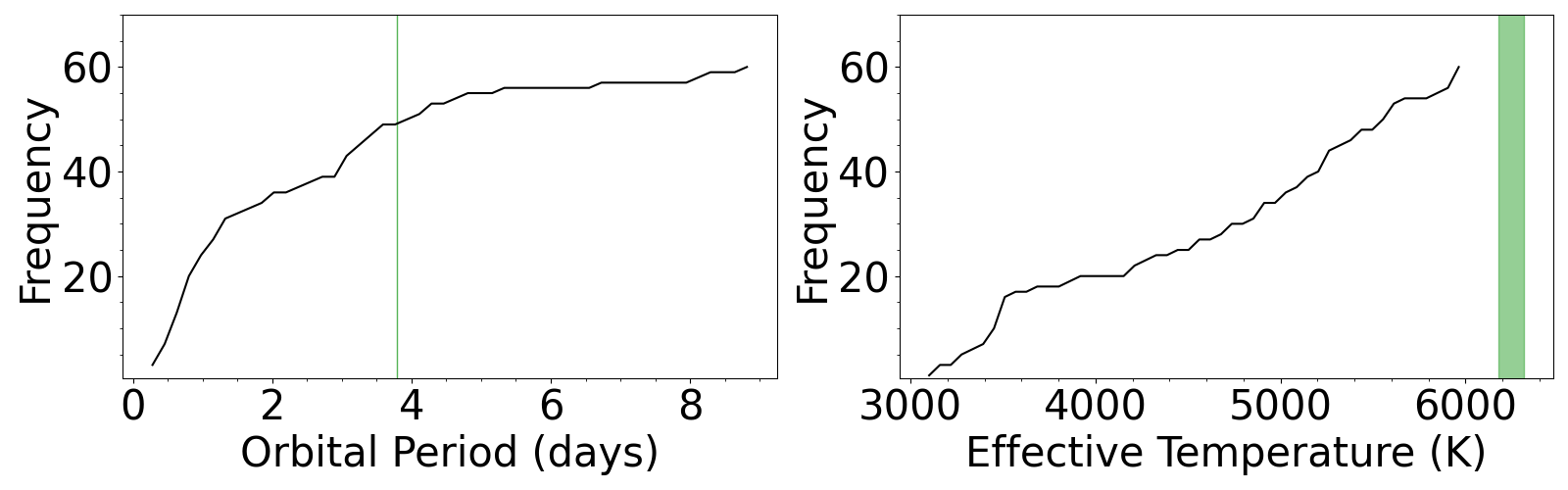}
    \caption{Cumulative distributions for the 60 transiting ND planets, displaying their masses, radii, orbital periods, and host star's effective temperatures in clockwise order from the top left. Vertical green bars highlight \Nplanet\, parameters, where the bar width represents their respective uncertainties.}
    \label{fig:cumulativePlots}
\end{figure}

Due to its impressively high yield of detected transiting planets, the Kepler Space Telescope \citep{borucki2010kepler} has played a major role in mapping out the various populations of worlds in inner planetary systems. From HJs to Earth-like planets, the mission provided a good overview of the galactic planetary landscape, for example showcasing that super-Earths and sub-Neptunes orbit $\sim$30$\%$ of Sun-like stars \citep{fressin2013false}. In addition, the detection of several Ultra Short Period (USP; P $\leq$ 1 day) planets \citep{sanchis2013transits,rowe2014validation}, the Neptune desert (R$_{\ast}$ 2-10 < R$_{\oplus}$ and P < 4 days), and the regularity of multi-planet systems \citep[e.g.,][]{lissauer2012almost,rowe2014validation} were amongst the highlights. Subsequent missions like TESS \citep{ricker2015transiting} not only contributed to confirming the Kepler milestone discoveries, but also expanded the population of all planetary types, including those highly sought-after like ND planets (e.g., LTT9779b, \Nplanet\,, TOI-824b and TOI-849b). This mission also enabled endeavours in planetary and stellar science, such as the study of orbital decay \citep[e.g., WASP-12;][]{yee2019orbit}, the detection of young planets \citep[e.g., TOI-6442;][]{alves2025ngts}, and investigations into asteroseismology and Gyrochronology in stellar science \citep{nielsen2020tess,bouma2023empirical}.

The ND represents a region in the mass–period diagram that shows a surprising paucity of Neptune-sized planets. The mechanisms proposed to explain this scarcity often involve a combination of planetary migration, atmospheric stripping through photoevaporation, and tidal interactions such as RLO, though the precise evolutionary channels remain uncertain. Thus far nearly 60 well-studied ND planets have been detected with masses of up to $\sim$ 31 M$_\oplus$. The selected ND sample in this study comes from the TEPCat catalogue of well-studied planets, those with masses up to $\sim$ 31 M$_\oplus$ falling inside the \citet{mazeh2016dearth} boundaries, highlighted by the green triangular region in Fig. \ref{fig:cumulativePlots}. The same figure shows the ND population cumulative distribution highlighting \Nplanet\,as a green vertical bar, with the width representing the parameter uncertainties. Our analysis reveals a planet that is more massive and larger than 77$\%$ and 82$\%$ respectively, of the population, while its orbital distance places it 81.5$\%$ farther, i.e., $\sim$ 18.5 $\%$ of the population orbital periods are longer than \Nplanet's. More interesting is that \Nplanet\, is hosted by a F7V, one of the first Neptune-like planets orbiting such a hot star. In fact, not many ND are detected around hotter than G-type stars, this sharp drop-off in their occurrence may be connected to the intense UV/XUV radiation rapidly stripping away their atmospheres. Therefore, \Nplanet\,can provide insights into how hot and cool hosts may influence ND planet evolution.

\subsection{Planet Structure and Internal Composition}
\label{sec:PSIC}
\begin{figure}
	\includegraphics[width=\columnwidth]{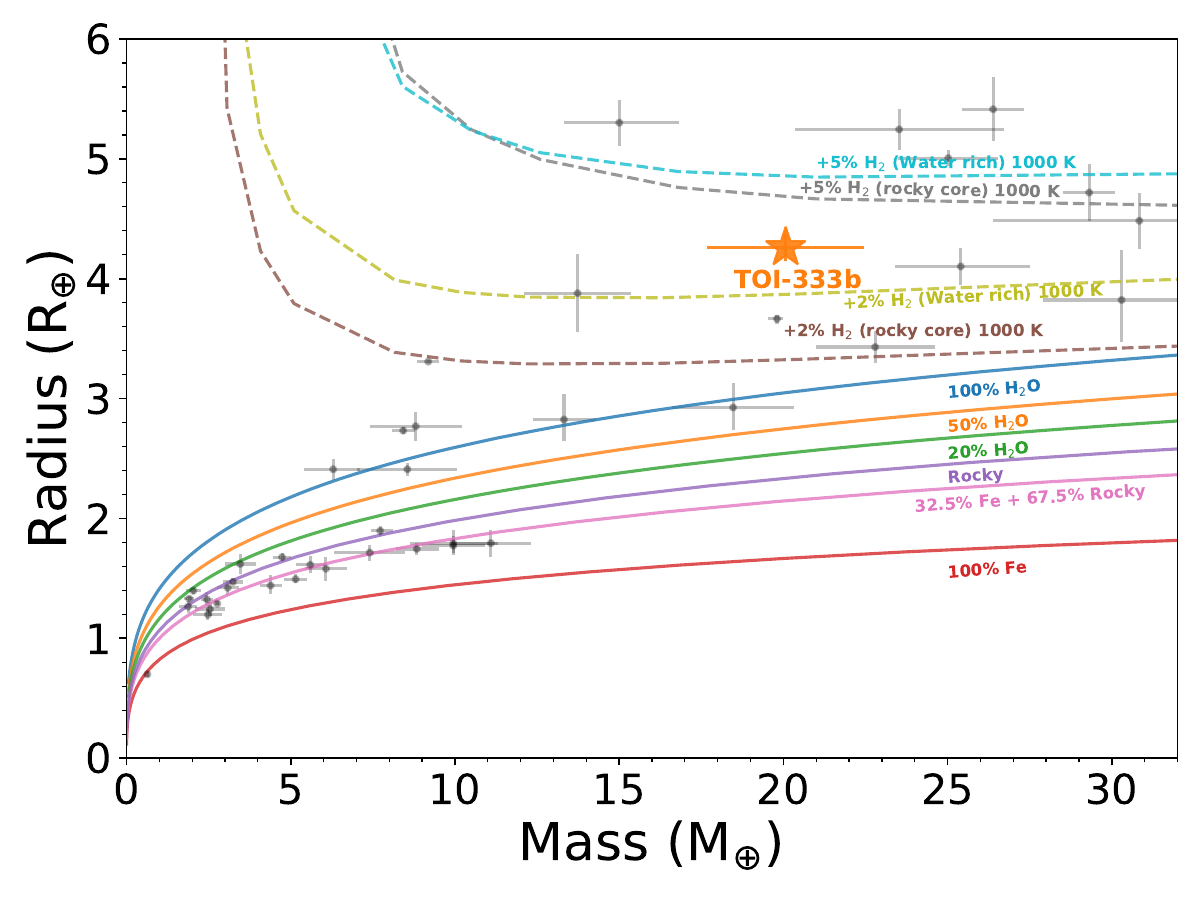}
    \caption{Radius as a function of mass for a set of well-studied ND planets from the TEPCat catalogue. The solid curves show compositional bi-layer models from \citealp{zeng2016mass}, ranging from 100$\%$ iron worlds to 100$\%$ water planets. Labels are colour-coded to represent the respective model.  Dashed-line models are taken from \citep{zeng2019growth}, showing 2$\%$ and 5$\%$ H$_2$ envelopes at 1000 K and distinct core compositions.}
    \label{fig:psic}
\end{figure}
Our analysis in $\S$ \ref{sub:globalmodelling} revealed a planet slightly larger than Neptune and with a similar mass, rendering a lower bulk density of 1.42 $\pm$ 0.21 \gccc. In Fig. \ref{fig:psic} we compare \Nplanet\,with composition models by \citet{zeng2016mass, zeng2019growth}. Solid lines represent two-layer models composed of H$_2$O, MgSiO$_3$ (rocky worlds) and/or Fe without atmospheres, while the dashed lines highlight the addition of an envelope with a gas temperature of 1,000 K. \Nplanet\, is represented by the orange star, and may have an H$_2$ envelope with a core composition ranging from either H$_2$O, MgSiO$_3$ or more likely a combination of both.

Using \texttt{smint}\footnote{https://github.com/cpiaulet/smint} \citep[Structure Model INTerpolator;][]{piaulet2021wasp}, we investigate the most suitable composition model. The code provides posterior distributions for the planet's core mass fraction (CMF) as well as the H/He envelope or H$_2$O mass fraction of a planet by interpolating over a set of composition models based on those from \citet{lopez2014understanding, zeng2016mass, aguichine2021mass}. \texttt{smint} requires planetary mass, radius, age, and insolation flux to perform the MCMC samplings. We opted for flat priors based on Tables \ref{tab:stellar} and \ref{tab:planet}, and a chain of 10,000, discarding 60$\%$ in the burn-in step, and employing 1,000 walkers.

Using \citet{aguichine2021mass} irradiated ocean world mass-radius relationships we found a 20$^{+11}_{-10}\%$ H$_2$O mass fraction with a core fraction of 35$^{+20}_{-23}\%$, representing equal parts of iron and silicate-based chemistry. Additionally, we used \texttt{smint} \citet{lopez2014understanding} models to estimate the gas-to-core mass ratio for a H/He envelope, yielding a fraction of 8.5$^{+10.9}_{-8.3}\%$. The lack of a significant H/He envelope may indicate that \Nplanet's internal composition is dominated by a rocky (MgSiO$_3$) composition with almost no H/He envelope or a water rich world. Nonetheless, we point out that further atmospheric follow-up is necessary to unveil the atmospheric chemistry, providing detailed insights into the planet's formation.
\subsection{Atmospheric follow-up with JWST}
The detection of transiting exoplanets has opened the door to studying their atmospheric compositions, with short-period planets continuing to be prime targets for atmospheric characterisation. Planets with high equilibrium temperatures (T${\rm eq} \geq 1000$ K) and low bulk densities ($\rho_{\rm p} < 0.3$ gcm$^{-3}$) tend to have extended atmospheres, allowing material from deeper layers to be probed via transmission spectroscopy \citep{seager2000theoretical}. In contrast, secondary eclipse observations make it possible to measure the planet’s day-side temperature and infer its albedo or reflectivity through emission spectroscopy. These techniques have been widely employed using both ground-based facilities (e.g., ESPRESSO/VLT, HIRES/Keck) and space-based observatories such as the HST, Spitzer, and more recently, JWST.

\begin{figure}
	\includegraphics[width=\columnwidth]{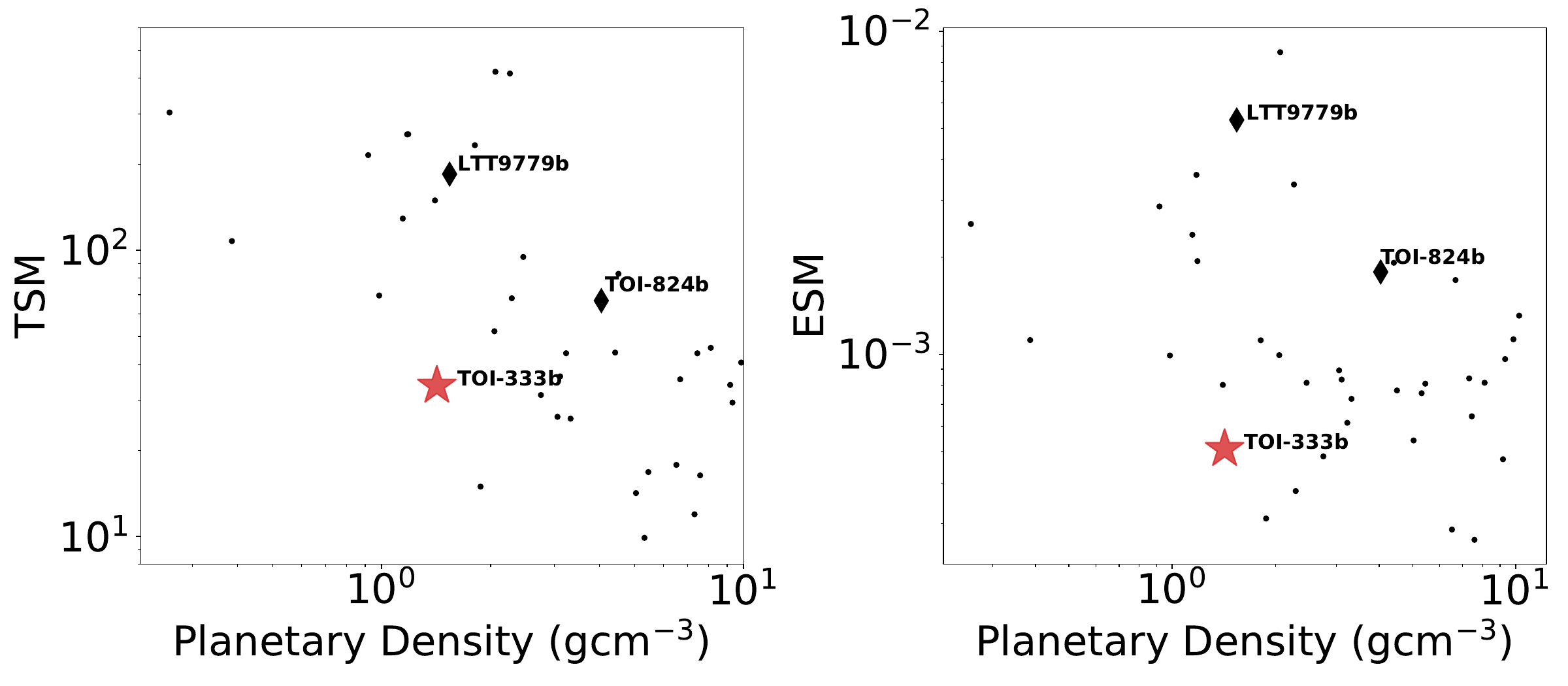}
    \includegraphics[width=\columnwidth]{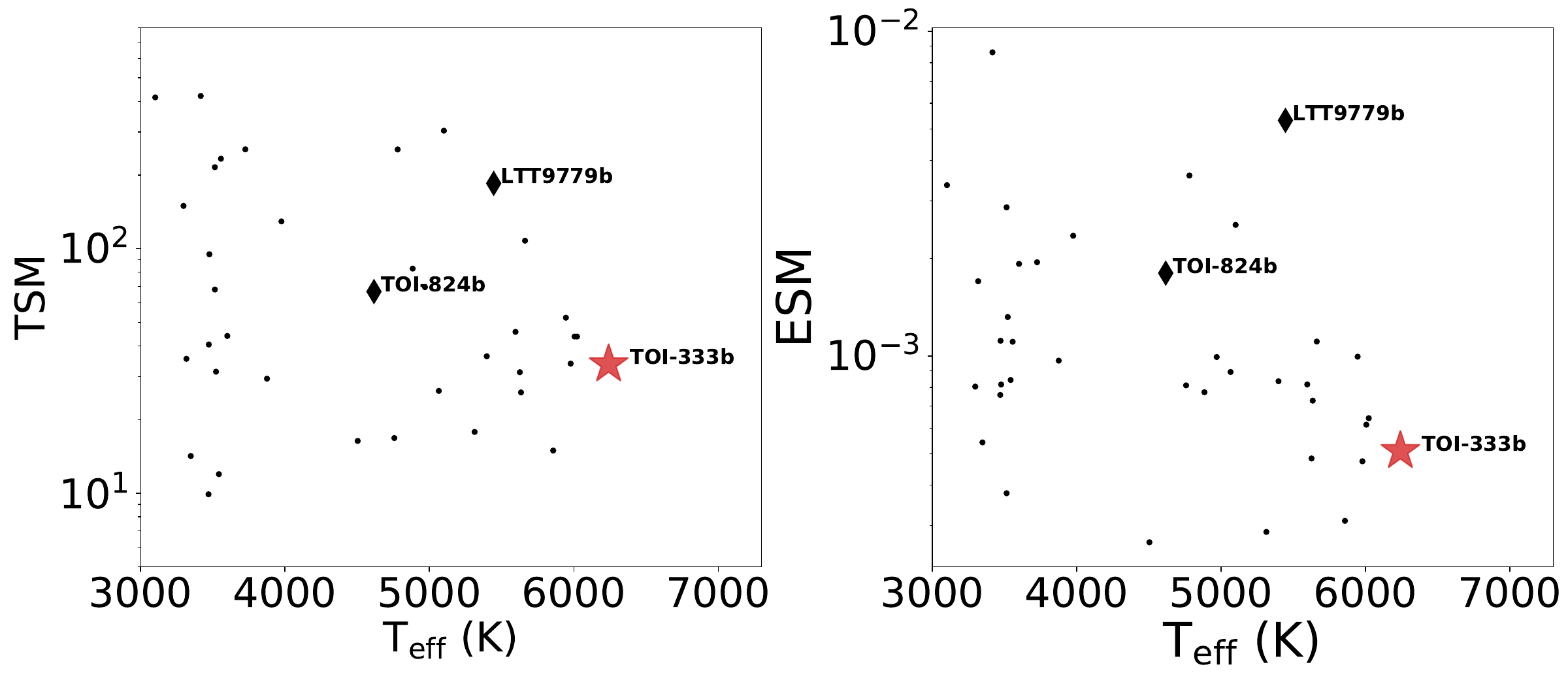}
    \caption{\textbf{Top}: Transmission and emission spectroscopy metrics as a function of planetary density for the transiting ND sample, shown as black circles. Black diamonds highlight two benchmark ND planets for comparison with TOI-333b, which is marked by a red star symbol. All TSM and ESM values were computed homogeneously using system parameters from the TEPCat catalogue.
    \textbf{Bottom}: TSM and ESM plotted against the host star’s effective temperature. The colour scheme and symbols follow those described in the top panel.}
    \label{fig:tsm-esm}
\end{figure}

Fig.~\ref{fig:tsm-esm} presents the transmission and emission spectroscopy metrics, dubbed TSM and ESM, respectively. They were computed homogeneously using Equations 1 and 4 from \citet{kempton2018framework}, for the ND population shown in Fig.~\ref{fig:per-rad-diag}. Benchmark ND planets LTT-9779b and TOI-824b are included for comparison. We note that \Nplanet\,exhibits a TSM value likely within JWST's capabilities, making it a compelling target for atmospheric follow-up. Notably, it lies at the hot end of the $T_{\rm eff}$–TSM parameter space, where very few ND hosts are present (T$_{\rm eff} > 6000$ K). Therefore, the detection of \Nplanet\, not only contributes to the scarce population of ND planets, (currently $\sim$60 known detections), but also offers a valuable opportunity to probe the atmospheric chemistry of ND planets in the hotter stellar regime—particularly given that its host star appears to exhibit low activity levels.

\section{Conclusions}
\label{sec:conc}

We report the discovery of \Nplanet, a ND planet with mass, radius, and bulk density of 20.1 $\pm$ 2.4 M$_{\oplus}$, 4.26 $\pm$ 0.11 R$_{\oplus}$, and 1.42 $\pm$ 0.21 \gccc, respectively. The planet orbits a F7V star every 3.78 d, whose mass, radius and effective temperature are 1.2 $\pm$ 0.1 \msun, 1.10 $\pm$ 0.03 \rsun, and 6241$^{+73}_{-62}$ K, respectively. \Nplanet\, is likely younger than 1 Gyr and older than a few hundred Myr, which is supported by the presence of the doublet Li line around 6707.856 $\AA$ and the comparison to Li abundances in open clusters with well constrained ages. Its host does not seem to present large photometric activity variation, which might hint for a pole-oriented orbit, which has been commonly observed for several ND planets, or a relatively quiet star. Finally, \Nplanet\,is more massive and larger than 77$\%$ and 82$\%$ of its population, respectively. Furthermore, its host star ranks among the hottest known for ND planets, making this system a unique laboratory to study the evolution of such planets around hot stars.

\section*{Acknowledgements}
DRA acknowledges support of ANID-PFCHA/Doctorado Nacional-21200343, Chile.  
JSJ greatfully acknowledges support by FONDECYT grant 1201371 and from the ANID BASAL project FB210003.
AS postdoctoral fellowship is funded by F.R.S.-FNRS research project ID 40028002 (Detection and Study of Rocky Worlds).
EG gratefully acknowledges support from UK Research and Innovation (UKRI) under the UK government’s Horizon Europe funding guarantee [grant number EP/Z000890/1].
ML acknowledges support of the Swiss National Science Foundation under grant number PCEFP2$\_$194576. The contribution of ML has been carried out within the framework of the NCCR PlanetS supported by the Swiss National Science Foundation under grant 51NF40$\_$205606.
DJA acknowledges this research was funded in part by the UKRI, (Grants ST/X001121/1, EP/X027562/1).
KAC acknowledges support from the TESS mission via subaward s3449 from MIT.
We acknowledge the use of public TESS data from pipelines at the TESS Science Office and at the TESS Science Processing Operations Center.
Resources supporting this work were provided by the NASA High-End Computing (HEC) Program through the NASA Advanced Supercomputing (NAS) Division at Ames Research Center for the production of the SPOC data products.
This work makes use of observations from the LCOGT network. Part of the LCOGT telescope time was granted by NOIRLab through the Mid-Scale Innovations Program (MSIP). MSIP is funded by NSF.
This research has also made use of the Exoplanet Follow-up Observation Program (ExoFOP; DOI: 10.26134/ExoFOP5) website, which is operated by the California Institute of Technology, under contract with the National Aeronautics and Space Administration under the Exoplanet Exploration Program.
Funding for the TESS mission is provided by NASA's Science Mission Directorate. 
The paper is also based on data collected under the NGTS project at the ESO La Silla Paranal Observatory. The NGTS facility is operated by the consortium institutes with support from the UK Science and Technology Facilities Council (STFC) under projects ST/M001962/1, ST/S002642/1 and ST/W003163/1.

\bibliographystyle{aa}
\bibliography{paper} 

\appendix
\section{EXTRA TABLES AND FIGURES}
\subsection{Photometric and RV time series}

\begin{figure*}
	\includegraphics[width=2\columnwidth,angle=0]{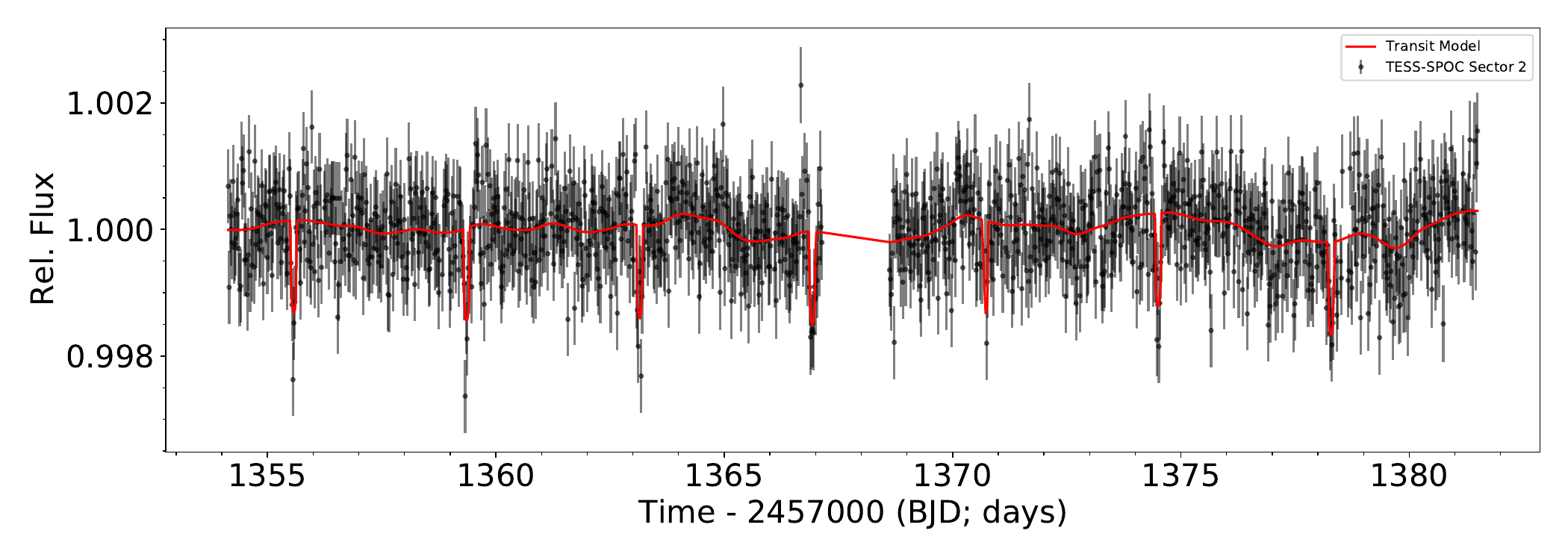}
    \includegraphics[width=2\columnwidth,angle=0]{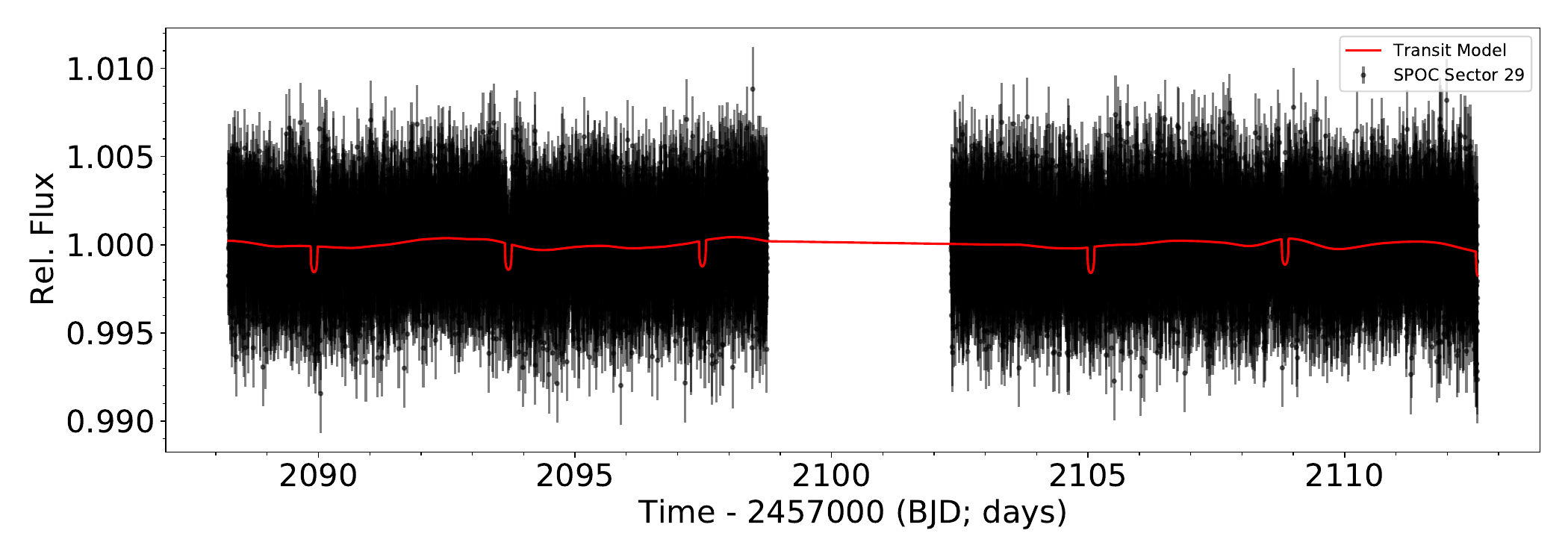}
    \includegraphics[width=2\columnwidth,angle=0]{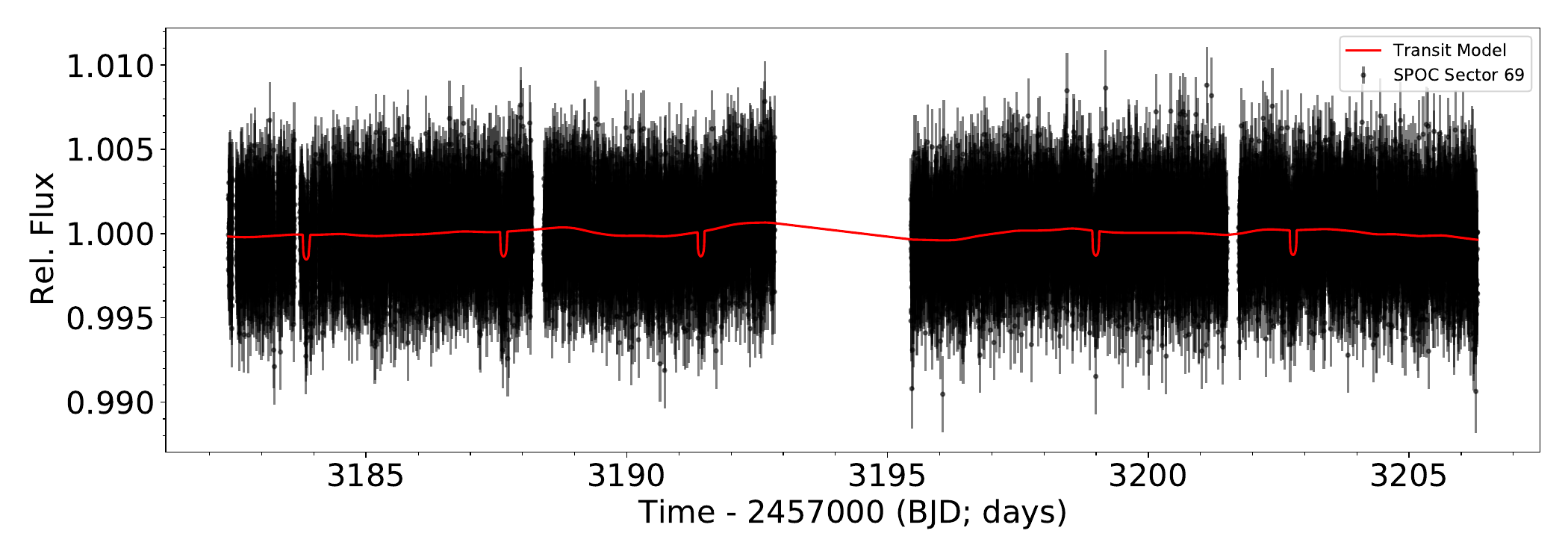}
    \includegraphics[width=2\columnwidth,angle=0]{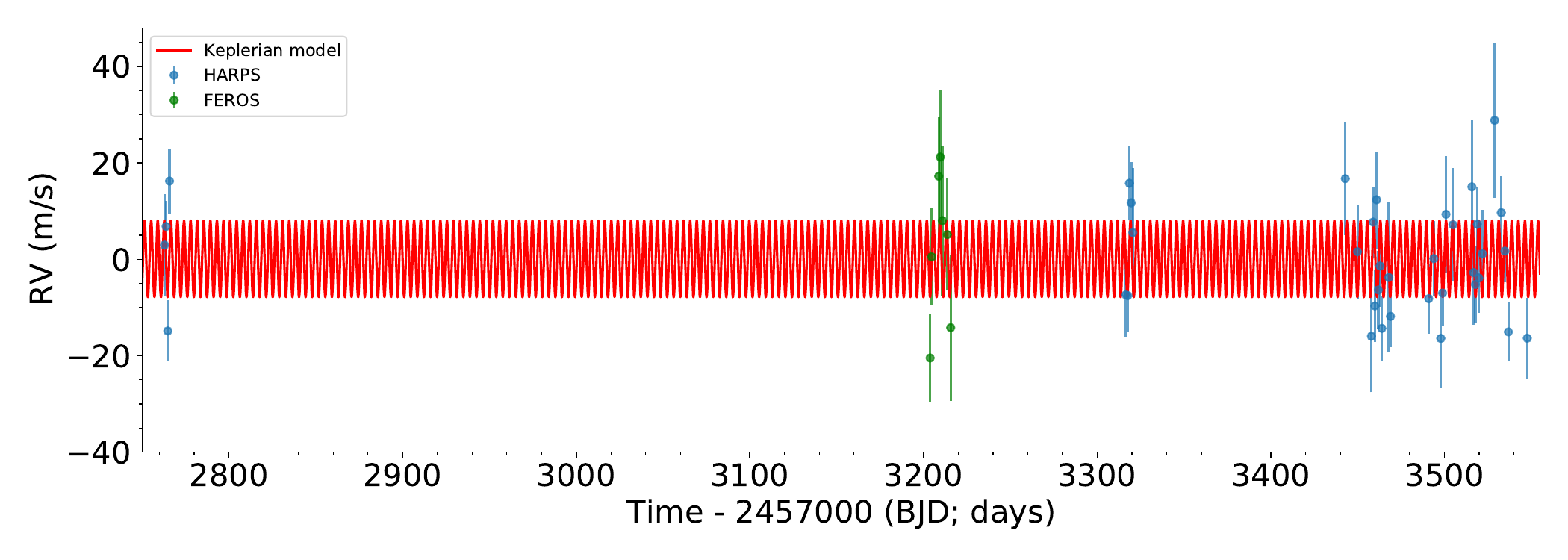}
    \caption{From top to bottom we show the TESS photometric time series with the best-fitting transit + GP model in red. Cadences are 30, 2, and 2 minutes, respectively. At the bottom, the RV time-series are highlighted with FEROS and HARPS instruments in light green and blue. The best-fitting Keplerian are shown in red. No GPs were used in the RVs part of the global model.}
    \label{fig:all-lc}
\end{figure*}

\begin{figure*}
	\includegraphics[width=2\columnwidth]{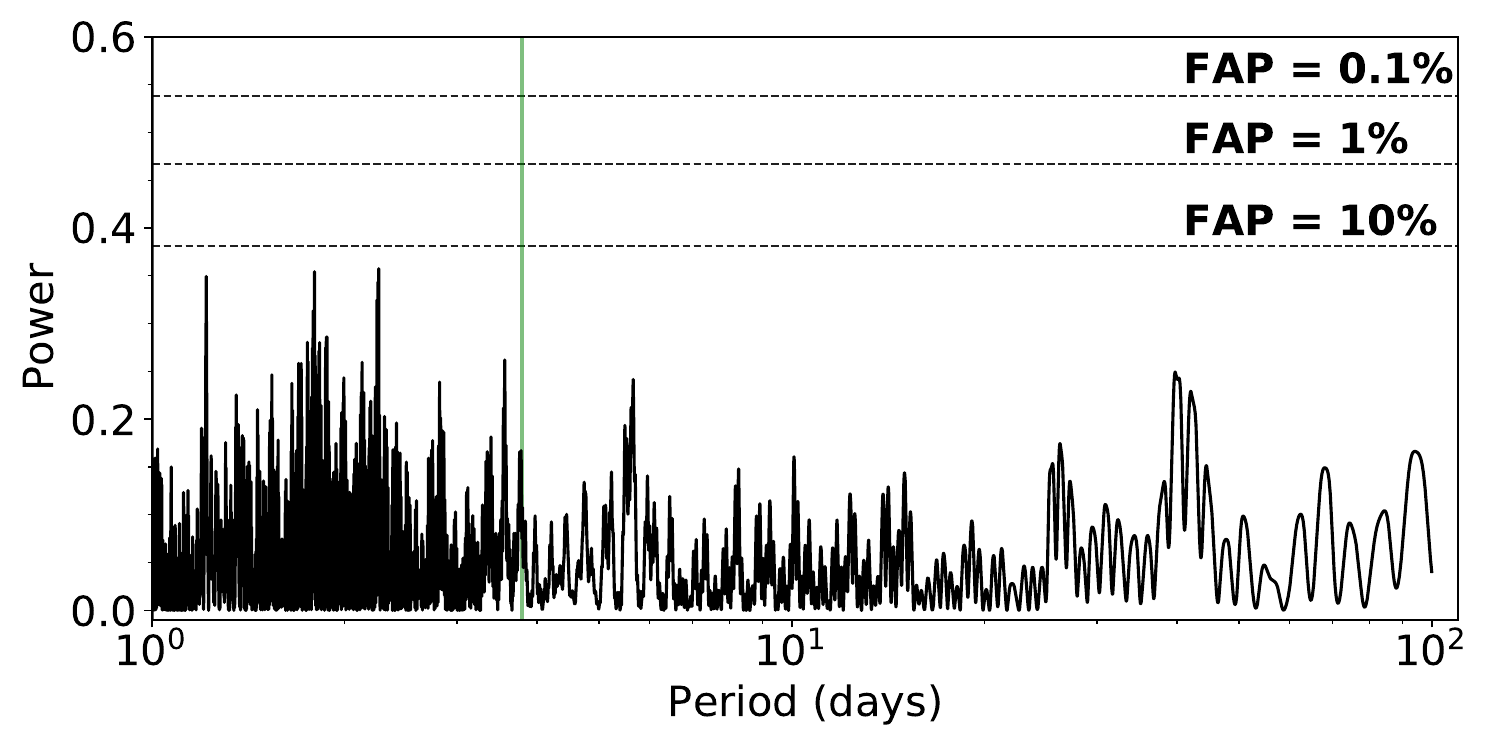}
    \caption{Lomb–Scargle periodogram on HARPS radial velocity residuals. Dashed black lines show the 0.1$\%$, 1$\%$, and 10$\%$ FAP levels (top to bottom). The green vertical bar marks the orbital period of TOI-333b.}
    \label{fig:RES-LS}
\end{figure*}

\subsection{Instrument's Photometric Time Series.}
\begin{table*}
	\centering
	\caption{TESS, LCOGT-SAAO, and NGTS photometry for \Nstar. The full table is available in a machine-readable format from the online journal.  A portion is shown here for guidance.}
	\label{tab:photometry}
	\begin{tabular}{cccc}
	Time	&	Flux       &Flux & Instrument\\
    (BJD$_{\rm TDB}$-2457000)	&	(normalised)	&error & \\
	\hline
    1354.3491055 & 1.0004 & 0.0006 & TESS \\
    1354.3699389 & 0.9997 & 0.0006 & TESS \\
    1354.3907724 & 0.9990 & 0.0006 & TESS \\
    1354.4116058 & 0.9991 & 0.0006 & TESS \\
    1354.4324393 & 1.0011 & 0.0006 & TESS \\
    ...  &   ...    &   ...  &  ...  \\
    1719.3266660 & 0.9985 & 0.0013 & LCOGT-SAAO \\
    1719.3273100 & 1.0022 & 0.0013 & LCOGT-SAAO \\
    1719.3279579 & 1.0013 & 0.0013 & LCOGT-SAAO \\
    1719.3286009 & 1.0011 & 0.0013 & LCOGT-SAAO \\
    1719.3292430 & 1.0000 & 0.0013 & LCOGT-SAAO \\
    1719.3298860 & 0.9986 & 0.0013 & LCOGT-SAAO \\
    ...  &   ...    &   ... & ...  \\
    3555.7743964 & 0.9981 & 0.0006 & NGTS \\
    3555.7757488 & 0.9989 & 0.0007 & NGTS \\
    3555.7771257 & 0.9963 & 0.0007 & NGTS \\
    3555.7785528 & 0.9991 & 0.0007 & NGTS \\
    3555.7799270 & 0.9995 & 0.0007 & NGTS \\
    3555.7812923 & 0.9975 & 0.0007 & NGTS \\
    3555.7827168 & 0.9992 & 0.0007 & NGTS \\
    3555.7840950 & 0.9976 & 0.0007 & NGTS \\
    3555.7854795 & 0.9987 & 0.0008 & NGTS \\
    3555.7868719 & 0.9978 & 0.0007 & NGTS \\
    3555.7882631 & 0.9980 & 0.0007 & NGTS \\
	\hline
	\end{tabular}
\end{table*}
\subsection{Speckle Imaging}
\begin{figure*}
	\includegraphics[width=2\columnwidth]{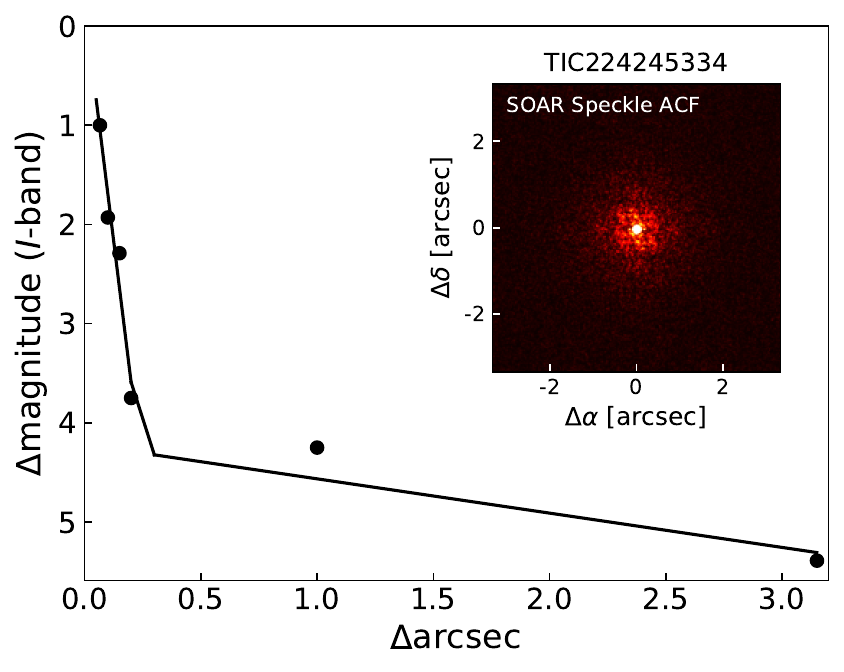}
    \caption{\Nstar\ speckle image obtained with the SOAR telescope showing no nearby binary was observed at 1 arcsec. }
    \label{fig:soarSpeckle}
\end{figure*}
\begin{figure*}
	\includegraphics[width=2\columnwidth]{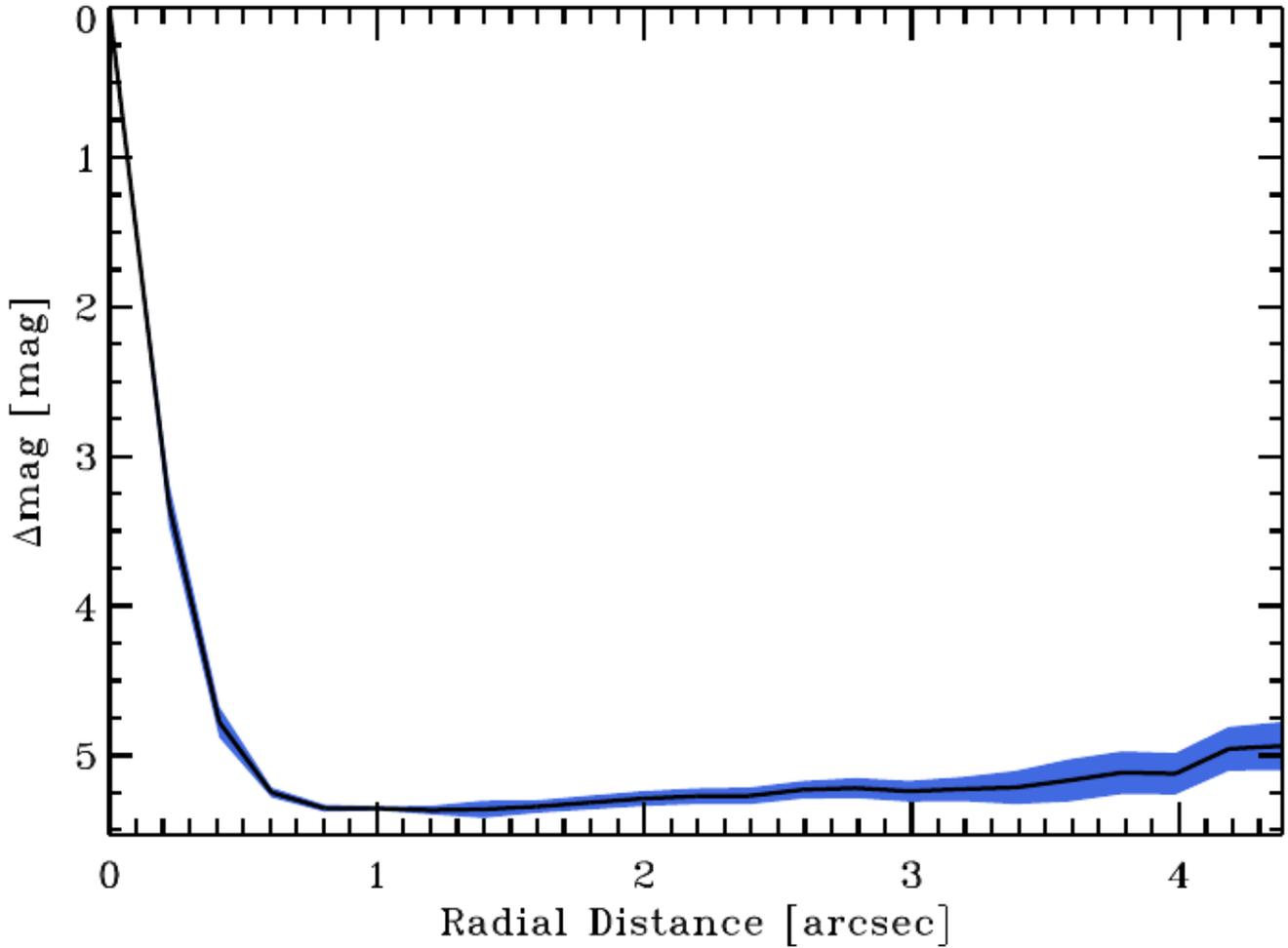}
    \caption{\Nstar\ speckle image obtained with the NaCo at the VLT telescope indicating that no companions were observed to $\Delta$mag$\sim$5.}
    \label{fig:NaCoAO}
\end{figure*}
\subsection{PHOENIX SED modelling}
\begin{figure*}
	\includegraphics[width=2\columnwidth]{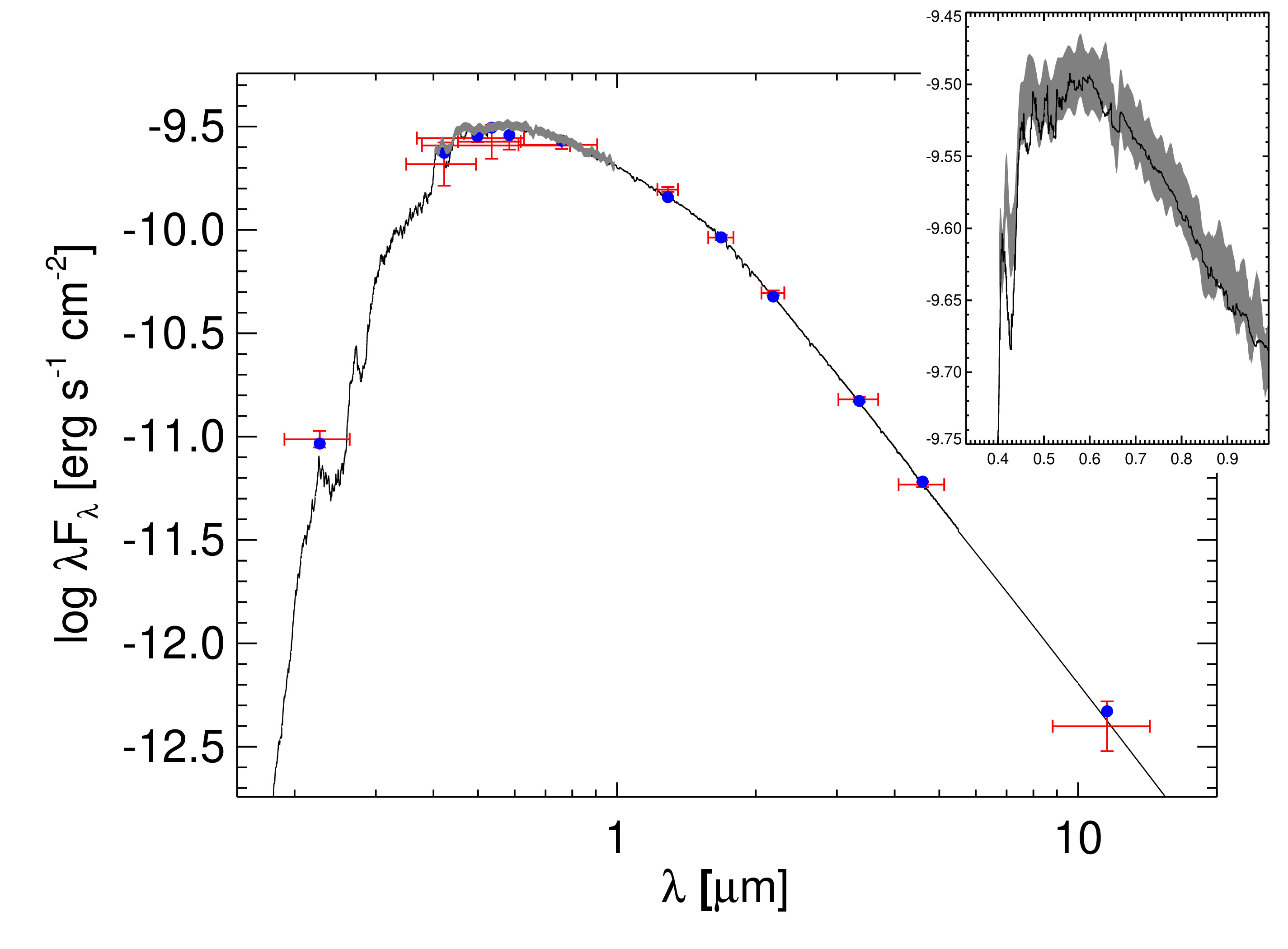}
    \caption{Spectral energy distribution of \Nstar. Red symbols represent the observed photometric measurements, where the horizontal bars represent the effective width of the passband. Blue symbols are the model fluxes from the best-fit PHOENIX atmosphere model (black). The absolute flux-calibrated {\it Gaia \/} spectrophotometry is shown as the grey swathe; see also the inset plot.
    \label{fig:phoenix-SED}}
\end{figure*}
\subsection{Lithium Equivalent Width}

\begin{figure*}
	\includegraphics[width=2\columnwidth]{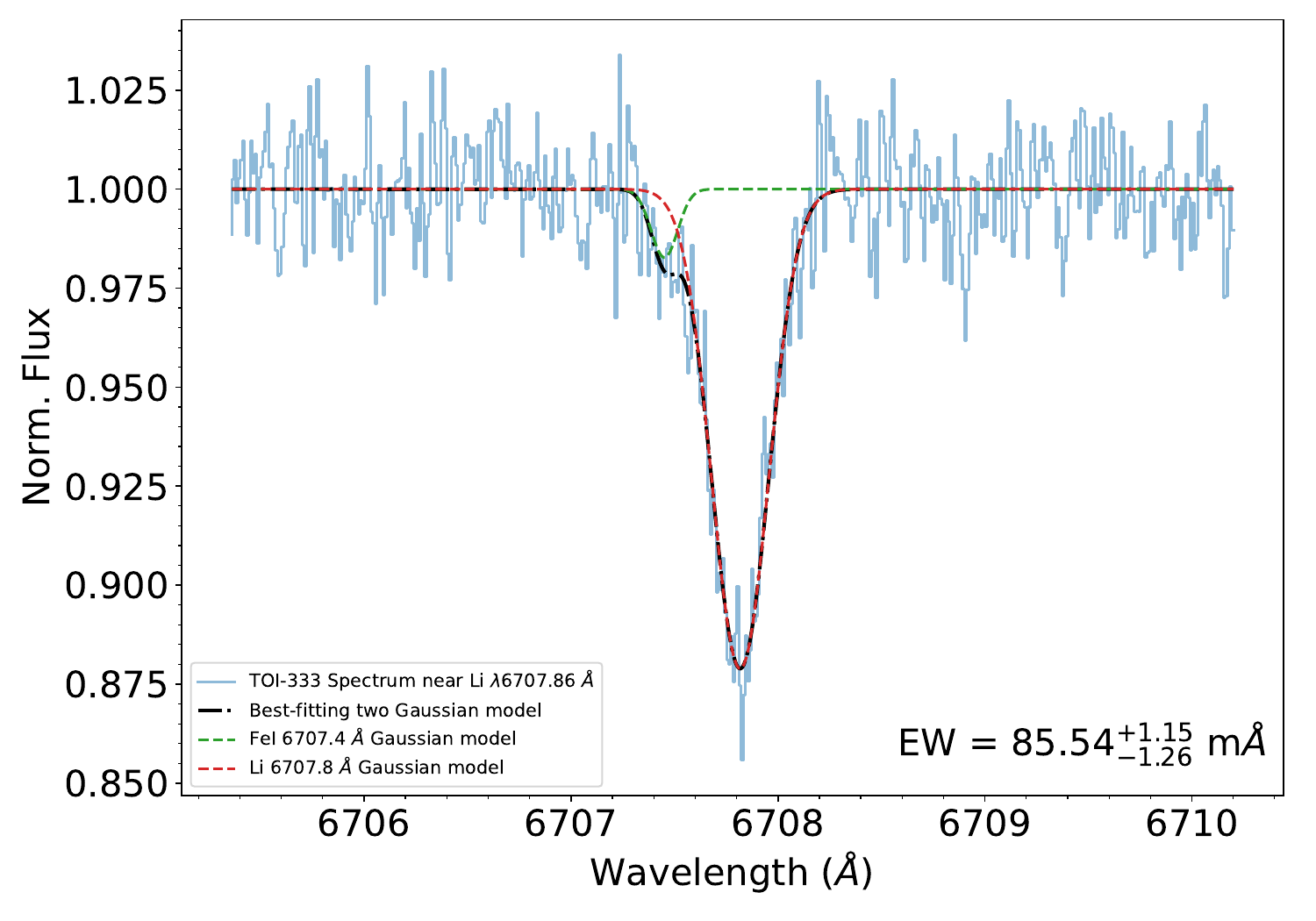}
    \caption{\Nstar~coadded spectra around the Lithium line at 6707.86 \AA\,in blue with a best-fitting double Gaussian model to the Fe (in green) and Li (in red) lines  Equivalent width. The total Gaussian model is shown in black.}
    \label{fig:Lithium}
\end{figure*}
\begin{figure*}
	\includegraphics[width=2\columnwidth]{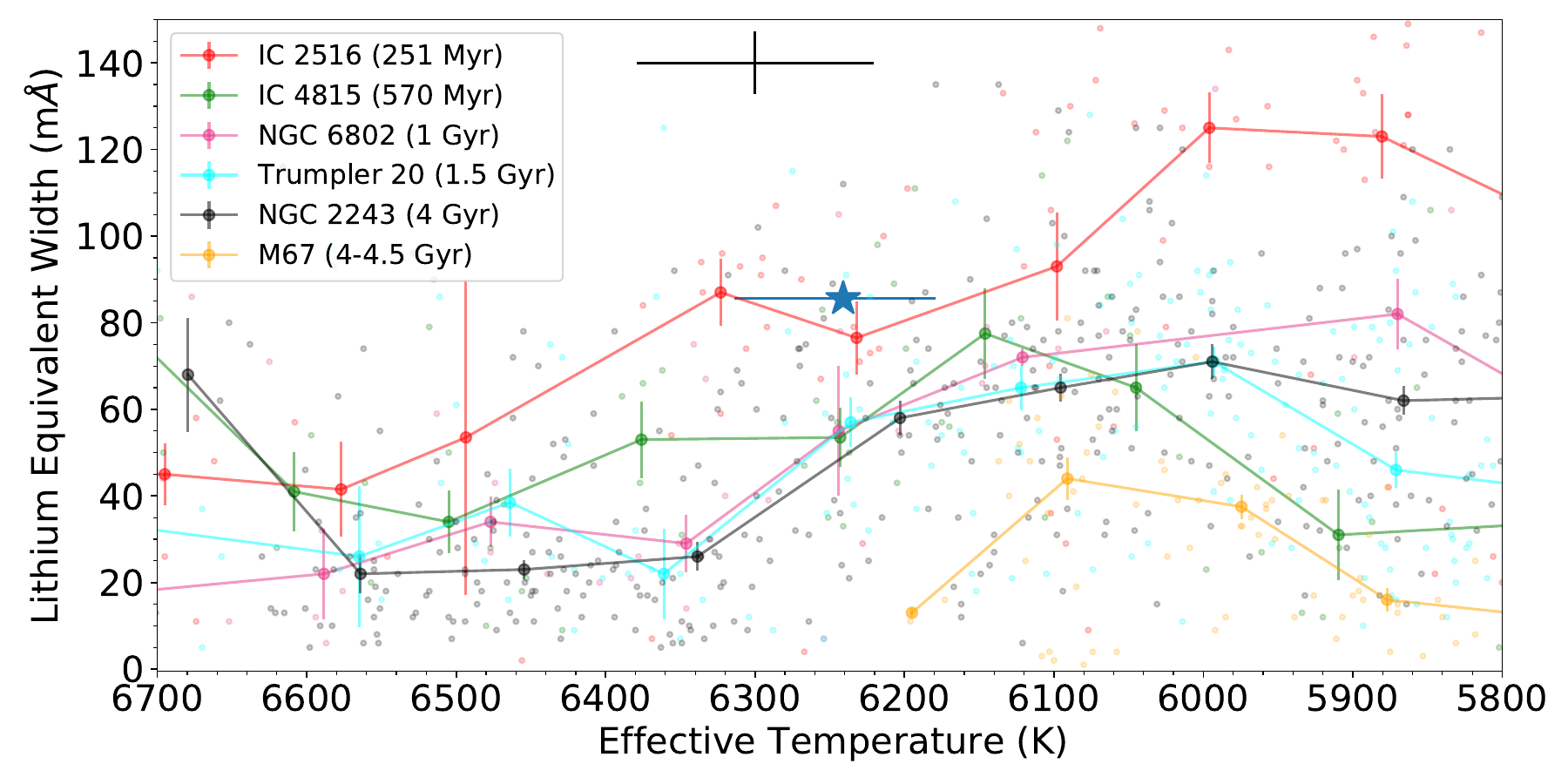}
    \caption{Open clusters Li EW $\lambda$\, 6707.86\AA\, as a function of effective temperature; the data are from \citep[][]{albarran2020gaia} colour coded by cluster's age. Circles with the same colour scheme shows the EW median binned for every 100 T$_{\rm eff}$ bins. \Nstar\,is represented by the blue star, while the black cross at the top shows the data typical 1-$\sigma$ confidence intervals.}
    \label{fig:ew-teff}
\end{figure*}

\subsection{HARPS AND FEROS Radial-Velocities}
\begin{table*}
	\centering
	\caption{HARPS and FEROS \Nstar~velocimetry obtained with the HARPS and FEROS spectrographs.}
	\label{tab:RVs-table}
	\begin{tabular}{cccc}
	Time	&	Radial Velocity       &Radial Velocity Error & Instrument\\
    (BJD$_{\rm TDB}$-2457000)	&	(m/s)	&(m/s) & \\
	\hline
2763.8164731	&303.3	&5.3	&HARPS\\
2764.7683173	&281.6	&6.4	&HARPS\\
2765.7712404	&312.7	&6.7	&HARPS\\
...                 & ...               & ...               & ...\\
3203.78669812      &    257.3          &9.10                &FEROS\\
3203.69766970      &    278.3          &10.00               &FEROS\\
3213.68682803      &    282.9          &11.60               &FEROS\\
3215.64779999      &    263.6          &15.20               & FEROS\\
...                 & ...               & ...               & ... \\
3316.5706793	&289.1	&8.7    &HARPS\\
3317.5713573	&288.9	&7.4    &HARPS\\
3521.8347154	&297.6	&9.1    &HARPS\\
3532.6913223	&306.1	&7.6    &HARPS\\
	\hline
	\end{tabular}
\end{table*}
\subsection{Transit Timing, Duration and Depth Variations}
\begin{table*}
	\centering
	\caption{\Nplanet~Transit Timing, Duration and Depth Variations}
	\label{tab:ttvs}
	\begin{tabular}{ccccc}
	Mid-Transit Times	&	TTVs       & TDV & Depth & Instrument\\
    (BJD$_{\rm TDB}$-2457000)	&	(minutes)	&(hour) & \\
1355.563497 & -10.66$_{-11.97}^{+14.63}$ & 3.06 $\pm$ 0.18 & 0.035 $\pm$ 0.003 & TESS\\
1359.352666 & -5.93$_{-19.66}^{+16.53}$ & 3.20 $\pm$ 0.24 & 0.035 $\pm$ 0.004 & TESS\\
1363.144010 & 3.22$_{-12.52}^{+17.49}$ & 3.38 $\pm$ 0.20 & 0.035 $\pm$ 0.003 & TESS\\
1366.932347 & 7.33$_{-15.97}^{+15.56}$ & 3.30 $\pm$ 0.24 & 0.035 $\pm$ 0.003 & TESS\\
1374.498032 & 0.82$_{-16.86}^{+19.75}$ & 3.27 $\pm$ 0.24 & 0.0345 $\pm$ 0.003 & TESS\\
1378.277460 & -8.25$_{-11.92}^{+12.70}$ & 3.17 $\pm$ 0.22 & 0.032 $\pm$ 0.003 & TESS\\
2089.939186 & 41.66$_{-19.96}^{+16.27}$ & 3.34 $\pm$ 0.22 & 0.035 $\pm$ 0.003 & TESS\\
2093.702461 & 10.00$_{-9.90}^{+9.61}$ & 3.08 $\pm$ 0.19 & 0.037 $\pm$ 0.004 & TESS\\
2108.816872 & -28.38$_{-19.47}^{+18.03}$ & 3.27 $\pm$ 0.25 & 0.030 $\pm$ 0.003 & TESS\\
3183.861055 & 20.08$_{-22.30}^{+35.20}$ & 3.26 $\pm$ 0.24 & 0.035 $\pm$ 0.004 & TESS\\
3187.635435 & 2.56$_{-19.99}^{+14.93}$ & 3.22 $\pm$ 0.23 & 0.036 $\pm$ 0.003 & TESS\\
3191.424700 & 8.93$_{-21.73}^{+21.34}$ & 3.23 $\pm$ 0.24 & 0.032 $\pm$ 0.003 & TESS\\
3198.985559 & -4.56$_{-11.78}^{+11.73}$ & 3.26 $\pm$ 0.22 & 0.035 $\pm$ 0.004 & TESS\\
3202.753351 & -29.21$_{-19.30}^{+23.45}$ & 3.26 $\pm$ 0.25 & 0.034 $\pm$ 0.004 & TESS\\
2468.440882 & 8.28$^{+4.51}{-4.44}$ & 3.21$\pm$ 0.01 & 0.035$\pm$0.001 & LCOGT-SAAO\\
3573.746391 & 24.60$^{+8.94}{-8.33}$ & (fixed) & (fixed) & NGTS\\
3611.593435 & 16.56$^{+9.19}{-8.46}$ & (fixed) & (fixed) & NGTS\\
	\hline
	\end{tabular}
\end{table*}
\subsection{\texttt{ARIADNE} priors for the stellar characterisation}
\begin{table*}
	\centering
	\caption{TOI-333 priors used in \texttt{ARIADNE}}
	\label{tab:priors-ariadne}
	\tabcolsep=0.11cm
	\begin{tabular}{cc} 
Parameters		&	Prior distribution	\\
\hline
\teff       &   $\mathcal{N}$(6267,$100^2$)\\
\logg       &   $\mathcal{N}$(4.42,$0.1^2$)\\
\met        &   $\mathcal{N}$(0.0,$0.05^2$)\\
Distance    &   $\mathcal{N}$(347,$10^2$)\\
\rstar      &   $\mathcal{N}$(1.1,$0.5^2$)\\
A$_{\rm V}$ &   $\mathcal{U}$(0.0,0.5)\\
\hline
	\end{tabular}
\end{table*}
\subsection{\Nstar\ chemical abundances from \texttt{SPECIES}}
\begin{table*}
	\centering
	\caption{\Nstar{} chemical abundaces from \texttt{SPECIES}}
	\label{tab:CheAb}
	\tabcolsep=0.11cm
	\begin{tabular}{ccc} 
Parameters		&	$\mu \pm \sigma$ & number of lines	\\
\hline
AlI       &   $-0.35 \pm 0.20$ & 1\\
CaI       &   $-0.18 \pm 0.12$ & 3\\
CrI       &   $-0.05 \pm 0.08$ & 7\\
CuI       &   $-0.19 \pm 0.20$ & 1\\
FeI       &   $0.11  \pm 0.07$ & 9\\
FeII      &   $-0.02 \pm 0.07$ & 8\\
MgI       &   $0.01  \pm 0.14$ & 2\\
MnI       &   $0.17  \pm 0.12$ &  3\\
NaI       &   $0.01  \pm 0.12$ &  3\\
NaI       &   $0.01  \pm 0.12$ &  3\\
NiI       &   $-0.06 \pm 0.10$ &  5\\
SiI       &   $0.17  \pm 0.10$ &  4\\
TiI       &   $-0.02 \pm 0.12$ &  3\\
TiII      &   $0.10  \pm 0.12$ &  3\\
\hline
	\end{tabular}
\end{table*}

\end{document}